\documentclass[11pt, letterpaper]{article}

\usepackage[letterpaper, margin=.75in]{geometry}
\usepackage{float}
\usepackage{subcaption}
\usepackage{makecell}
\usepackage{authblk}
\usepackage{multirow}
\usepackage{hyperref}
\usepackage{graphicx}
\usepackage{amsfonts,amsmath,amssymb,amsthm}
\usepackage{stackengine}
\newcommand{\para}[1]{\vspace*{.15cm}\noindent{\textbf{#1}}}

\newtheorem{theorem}{Theorem}[section]
\newtheorem{lemma}[theorem]{Lemma}
\newtheorem{corollary}[theorem]{Corollary}
\newtheorem{definition}[theorem]{Definition}

\title{Computing Threshold Circuits with Void Reactions in Step Chemical Reaction Networks\thanks{This research was supported in part by National Science Foundation Grant CCF-2329918.}} 

\author[1]{Rachel Anderson}

\author[1]{Alberto Avila}

\author[1]{Bin Fu}

\author[2]{Timothy Gomez}

\author[1]{Elise Grizzell}

\author[1]{Aiden Massie}

\author[1]{Gourab Mukhopadhyay}

\author[1]{Adrian Salinas}

\author[1]{Robert Schweller}

\author[3]{Evan Tomai}

\author[1]{Tim Wylie}

\affil[1]{University of Texas Rio Grande Valley} 

\affil[2]{
Massachusetts Institute of Technology}
\affil[3]{
University of Texas Dallas}

\date{}

\begin{document}

\maketitle

\begin{abstract}
We introduce a new model of \emph{step} Chemical Reaction Networks (step CRNs), motivated by the step-wise addition of materials in standard lab procedures. Step CRNs have ordered reactants that transform into products via reaction rules over a series of steps. We study an important subset of weak reaction rules, \emph{void} rules, in which chemical species may only be deleted but never changed. We demonstrate the capabilities of these simple limited systems to simulate threshold circuits and compute functions using various configurations of rule sizes and step constructions, and prove that without steps, void rules are incapable of these computations, which further motivates the step model. Additionally, we prove the coNP-completeness of verifying if a given step CRN computes a function, holding even for $O(1)$ step systems. 


\end{abstract}


\section{Introduction}
\vspace{-.2cm}

Chemical Reaction Networks (CRNs) are one of the most established and longest studied models of self-assembly \cite{Aris:1965:ARMA, Aris:1968:ARMA}. CRNs originate in attempting to model chemical interactions as molecular species that react and create products from the reaction. This can be represented as an original number of each species and a set of replacement rules.
The fundamental nature of the model is evident in the independent inception of equivalent models in multiple areas of research through other motivations \cite{Cook:2009:AB}, such as Vector Addition Systems (VASs) \cite{Karp:1969:JCSS} and Petri-Nets \cite{Petri:1962:PHD}. Further, Population Protocols \cite{Angluin:2006:DC} are a restricted version where the number of input and output elements are each two.

\para{Step CRNs.} We propose and investigate an important but straightforward extension to the CRN model motivated by the desire to reflect standard laboratory and medical practices.  The \emph{Step} CRN models augments the CRN model with a sequence of discrete steps where an additional specified amount of chemical species is combined with the existing CRN after running the system to completion.  Our goal is to explore the computational power of Step CRNs using highly restricted classes of CRN rules that would otherwise be computationally weak.  In particular, we consider the problem of implementing the computationally universal class of \emph{Threshold Circuits} using only \emph{void} rules.



\para{Void Rules.} We study the computational power of Step CRNs under an extremely simple subset of CRN rules termed \emph{void} rules~\cite{Alaniz:2022:ARXIV}.  General CRN rules are powerful since they allow the removal, addition, and replacement of species.  Impressively, these rules have successful experimental implementations using DNA strand replacement mechanisms~\cite{soloveichik2010dna}.  However, implementing this level of generality requires sophisticated, and large, DNA complexes that incur practical errors and constitute one of the primary hurdles limiting the scalable implementation of molecular computing schemes~\cite{cardelli2020electric,wang2018effective}.  In contrast, void rules only have the ability to delete species.  Such a simple subset of reactions could plausibly permit drastically simpler and scalable molecular implementations based solely on the pair-wise bonding strength of single-stranded DNA. The only drawback is the inability of void rule systems to compute even simple functions.  We show that void rules become computationally powerful in the step model with just tri-molecular or bi-molecular interactions.  Specifically, we show how Threshold Circuits ($TC$), a powerful class of circuits with applications in deep learning, are simulated with void rules using a number of steps linear in the circuit's depth.  Our utilization of steps under this void rule restriction is necessary, as we further show that even simple circuits require the use of steps when restricted to pure void rules.

\vspace{-.3cm}

\subsection{Previous Work}
\vspace{-.3cm}

\para{Computation in Chemical Reaction Networks.}
 Stochastic Chemical Reaction Networks are only Turing-complete with the possibility for error \cite{soloveichik2008computation} while error-free stochastic Chemical Reaction Networks can compute precisely the set of semilinear functions \cite{angluin2007computational, chen2014deterministic}. CRNs have also recently been shown to be experimentally viable through DNA Strand Displacement (DSD) systems \cite{soloveichik2010dna}, and several CRN to DSD compilers have been created \cite{badelt2017general, lakin2012design, thubagere2017compiler, zhang2011dynamic}. 

\para{Boolean Circuits.} 
Using molecules for information storage and Boolean logic is a deep field of study. Here, we show a few highlights, starting with one of the first discussions in 1988 \cite{Aviram_1988} and an initial presentation of circuits with CRNs in 1991 \cite{hjelmfelt1991chemical}. Since then, the area has been extensively studied in CRNs and related models \cite{arkin1994computational, beiki2018real, cardelli2018chemical, Cook:2009:AB, ellis2019robust, jiang2013digital, lin2020mining, qian2011scaling, Qian_Winfree_2011}. Numerous gates have been built experimentally and proposed theoretically such as the AND \cite{cardelli2018chemical, dalchau2015probabilistic, magri2009fluorescent, Qian_Winfree_2011, thachuk2015leakless, Xiao_Zhang_Zhang_Chen_Shi_2020}, OR \cite{cardelli2018chemical, dalchau2015probabilistic, Qian_Winfree_2011, thachuk2015leakless}, NOT \cite{cardelli2018chemical}, XOR \cite{cardelli2018chemical, Xiao_Zhang_Zhang_Chen_Shi_2020}, NAND \cite{cardelli2018chemical, Cook:2009:AB, ellis2019robust, winfree2019chemical}, NOR \cite{cardelli2018chemical}, Parity \cite{eshra2013odd, Fan_Fan_Wang_Dong_2018, fan2022engineering}, and Majority  \cite{angluin2008simple, cardelli2012cell, Mailloux_Guz_Zakharchenko_Minko_Katz_2014}. 
Symmetric boolean functions of $n$ variables such as Majority have been found to have a circuit depth of $O(\log n)$ when implemented by AND, OR and NOT gates \cite{sergeev2014upper}.


\para{Void Rules.} The reachability problem, with systems of only void rules in proper CRNs, was studied in \cite{Alaniz:2022:ARXIV}. Previous studies had included void rules as a part of their systems but were never studied exclusively. The CRN++ programming language \cite{vasic2020crn++} allows these reactions to be programmed using the \textit{sub} module. However, if any product(s) remain, they can differ from the reactants. They can also be considered a subcategory of the broader concept of the extinction of rules and species in a system as referred to in \cite{winfree2019chemical}.     

\para{Mixing Systems.} Another generalization of CRNs that is closely related to the step model is I/O CRNs \cite{ellis2019robust}, where additional inputs can be added at timed intervals. Still, those inputs are read-only in the system (used exclusively as catalysts). Step CRNs generalize I/O CRNs as the inputs are not read-only and are rate-independent, unlike I/O CRNs.  
Staged systems have been explored in many self-assembly models\cite{chalk2018optimal, cirlos2023simulation, demaine2013one, mo2013iterative}. 

\vspace{-.4cm}

\subsection{Our Contributions}
\vspace{-.2cm}
Table \ref{tab:results} has an overview of the main results of this paper beyond the introduction of the model and simulation definitions. The most important results being the ability to simulate the class of Threshold Circuits (TC) by simulating AND, OR, NOT, and MAJORITY gates through a restrictive definition of simulation while only using small void rules. 

In Section \ref{sec:prelim}, we define Step Chemical Reaction Networks and what it means to compute a function. Following, in Section \ref{sec:3_0_circuits}, we show how to simulate the class TC of Threshold Circuits with void rules of size $(3,0)$ (rules where $3$ species react to delete each other) using $O(D \log f)$ steps, where $D$ is the depth of the circuit and $f$ denotes the maximum fan-out of the circuit. In Section \ref{sec:2_0_formulas}, we achieve the same result using both $(2,0)$ and $(2,1)$ rules and a slightly more efficient step complexity of $O(D)$. We then use exclusively $(2,1)$ rules to achieve this same result by adding a factor of $\log F_{maj}$ to the steps, where $F_{maj}$ is the maximum fan-in of majority gates. In Section \ref{sec:LBs}, we show there exist functions that require a logarithmic number of steps when restricted to constant reaction size, as well as the existence of $O(1)$-depth threshold circuits of fan-out $f$ that require $\Omega(\log f)$ steps, which matches the $O(D \log f)$ upper bound for $(3,0)$ circuits.  Finally, we show that it is coNP-complete to know whether a function can be strictly simulated by a step CRN system.


\begin{table}[t]
\vspace{-.2cm}
    \centering
\begin{tabular}{| c | c | c  | c | c | c |}\hline
\multicolumn{6}{|c|}{\textbf{Function Computation}}\\ \hline\hline
 \textbf{Rules} & \textbf{Species} & \textbf{Steps} & \textbf{Simulation} &\textbf{Family} & \textbf{Ref} \\ \hline
 $(3,0)$ & $O(\min(W^2, G F_{out}))$ & $O(D\log F_{out})$  & Strict & \textbf{TC} Circuits & Theorem~\ref{(3,0)_circuit} \\ \hline
 $(2,0) (2,1)$ & $O(G)$ & $O(D)$ & Strict & \textbf{TC} Circuits & Theorem~\ref{(2,0)(2,1)_circuit} \\ \hline
 $(2,1)$ & $O(G)$ & $O(D\log F_{maj})$ & Strict & \textbf{TC} Circuits & Corollary~\ref{(2,1)_circuit} \\ \hline
 
 \hline 
 
 $(c,0)$ & any & $\Omega(\log k)$ & Strict & $k$-CNOT & Theorem~\ref{thm:ccnotLB} \\
 \hline 

\end{tabular}

\vspace{.2cm}

\begin{tabular}{| c | c | c  | c | }\hline
\multicolumn{4}{|c|}{\textbf{Strict Function Verification}}\\ \hline\hline
 \textbf{Rules} & \textbf{Steps} & \textbf{Complexity} & \textbf{Ref} \\ \hline
 $(3,0)$ & $O(1)$ & coNP-complete & Theorem~\ref{coNP-c} \\ \hline
\end{tabular}
    \caption{Summary of $n$-bit circuit simulation results. $D$ is the depth of the circuit, $W$ is the width, $G$ is the number of gates in a circuit or number of operators in a formula, $F_{out}$ is the max fan-out, and $F_{maj}$ is the max fan-in of majority gates. \textbf{TC} stands for Threshold Circuits. The $k$-CNOT is a $k$ fan-in generalization of a Controlled NOT gate. Rule size $(c,0)$ means the row holds for all integer constants $c > 0$. 
    }
      \label{tab:results}
      \vspace{-.7cm}
\end{table}

\section{Preliminaries} \label{sec:prelim}

\vspace*{-.2cm}
\subsection{Chemical Reaction Networks}
\vspace{-.2cm}
\para{Basics.}
Let $\Lambda= \{\lambda_1, \lambda_2, \ldots, \lambda_{|\Lambda|}\}$ denote some ordered alphabet of \emph{species}. A configuration over $\Lambda$ is a length-$|\Lambda|$ vector of non-negative integers that denotes the number of copies of each present species. A \emph{rule} or \emph{reaction} is represented as an ordered pair of configuration vectors $R=(R_r, R_p)$. $R_r$ contains the minimum counts of each \emph{reactant} species necessary for reaction $R$ to occur, where reactant species are either \emph{consumed} by the rule in some count or leveraged as \emph{catalysts} (not consumed); in some cases a combination of the two. The \emph{product} vector $R_p$ has the count of each species \emph{produced} by the \emph{application} of rule $R$, effectively replacing vector $R_r$.  The species corresponding to the non-zero elements of $R_r$ and $R_p$ are termed \emph{reactants} and \emph{products} of $R$, respectively.

The \emph{application} vector of $R$ is $R_a = R_p - R_r$, which shows the net change in species counts after applying rule $R$ once. For a configuration $C$ and rule $R$, we say $R$ is applicable to $C$ if $C[i] \geq R_r[i]$ for all $1\leq i\leq |\Lambda|$, and we define the \emph{application} of $R$ to $C$ as the configuration $C' = C + R_a$. For a set of rules $\Gamma$, a configuration $C$, and rule $R\in \Gamma$ applicable to $C$ that produces $C' = C + R_a$, we say $C \rightarrow^1_\Gamma C'$, a relation denoting that $C$ can transition to $C'$ by way of a single rule application from $\Gamma$.
We further use the notation $C\rightarrow^*_\Gamma C'$ to signify the transitive closure of $\rightarrow^1_\Gamma$ and say $C'$ is \emph{reachable} from $C$ under $\Gamma$, i.e., $C'$ can be reached by applying a sequence of applicable rules from $\Gamma$ to initial configuration $C$. Here, we use the following notation to depict a rule $R=(R_r, R_p)$:
$ \sum_{i=1}^{|\Lambda|} R_r[i]s_i \rightarrow \sum_{i=1}^{|\Lambda|} R_p[i]s_i$.

For example, a rule turning two copies of species $H$ and one copy of species $O$ into one copy of species $W$ would be written as $2H + O \rightarrow W$.

\vspace{-.1cm}
\begin{definition}[Discrete Chemical Reaction Network]  A discrete chemical reaction network (CRN) is an ordered pair $(\Lambda, \Gamma)$ where $\Lambda$ is an ordered alphabet of species, and $\Gamma$ is a set of rules over $\Lambda$.
\vspace{-.1cm}
\end{definition}


An initial configuration and CRN $(\Lambda, \Gamma)$ is said to be \emph{bounded} if a terminal configuration is guaranteed to be reached within some finite number of rule applications starting from configuration $I$. We denote the set of reachable configurations of a CRN as $REACH_{I,\Lambda,\Gamma}$. A configuration is called \emph{terminal} with respect to a CRN $(\Lambda, \Gamma)$ if no rule $R$ can be applied to it. We define the subset of reachable configurations that are terminal as $TERM_{I,\Lambda,\Gamma}$. 


\vspace{-.2cm}
\subsection{Void Rules}
\begin{definition}[Void and Autogenesis rules]
A rule $R=(R_r, R_p)$ is a \emph{void} rule if $R_a = R_p - R_r$ has no positive entries and at least one negative entry.  A rule is an \emph{autogenesis} rule if $R_a$ has no negative values and at least one positive value.  If the reactants and products of a rule are each multisets, a void rule is a rule whose product multiset is a strict submultiset of the reactants, and an autogenesis rule one where the reactants are a strict submultiset of the products. There are two classes of void rules, catalytic and true void. In catalytic void rules, such as $(2,1)$ rules, one or more reactants remain, and one or more is deleted after the rule is applied. In true void rules, such as $(2,0)$ and $(3,0)$ rules, there are no products remaining.  

\end{definition}

\begin{definition}
The \emph{size/volume} of a configuration vector $C$ is $\verb"volume"(C) = \sum C[i]$.
\end{definition}

\begin{definition}[size-$(i,j)$ rules]
A rule $R=(R_r, R_p)$ is said to be a size-$(i,j)$ rule if $(i,j) = (\verb"volume"(R_r),$ $\verb"volume"(R_p))$. A reaction is trimolecular if $i=3$, bimolecular if $i = 2$, and unimolecular if $i = 1$.
\end{definition}

\vspace{-.3cm}
\subsection{Step CRNs} 
\begin{figure}[t]
    \vspace{-.2cm}
    \centering
    \includegraphics[width=1.\textwidth]{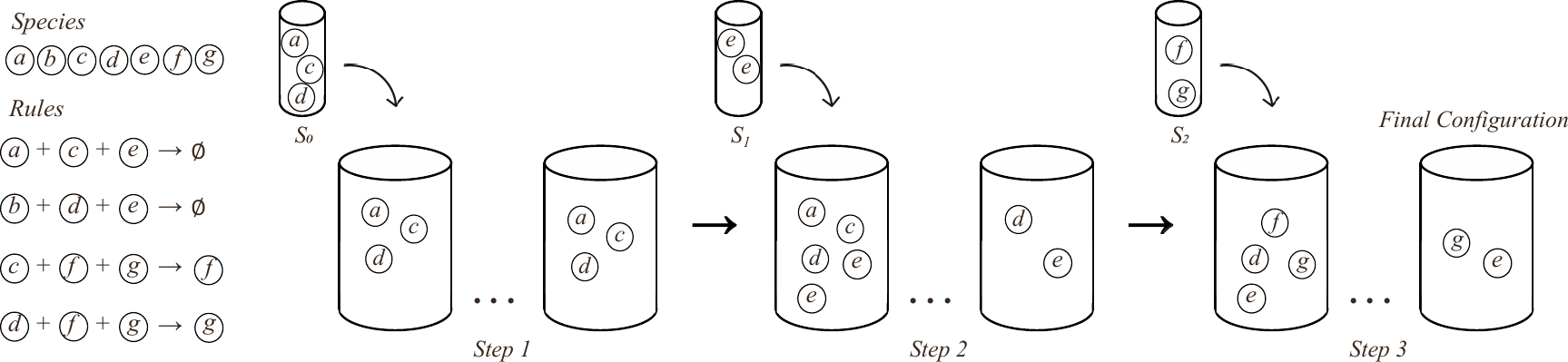}
    \vspace{-.2cm}
    \caption{An example step CRN system. The test tubes show the species added at each step and the system with those elements added. The CRN species and void rule-set are shown on the left.}
    \label{fig:step crn species and rules}
    \vspace{-.3cm}
\end{figure}

A step CRN is an augmentation of a basic CRN in which a sequence of additional copies of some system species are added after a terminal configuration is reached. Formally, a step CRN of $k$ steps is an ordered pair $( (\Lambda,\Gamma), (s_0, s_1, s_2, \ldots, s_{k-1}))$, where the first element of the pair is a normal CRN $(\Lambda,\Gamma)$, and the second is a sequence of length-$|\Lambda|$ vectors of non-negative integers denoting how many copies of each species type to add after each step. Figure \ref{fig:step crn species and rules} illustrates a simple step CRN system.

Given a step CRN, we define the set of reachable configurations after each sequential step.  Let $REACH_1$ be the set of reachable configurations of $(\Lambda, \Gamma)$ with initial configuration $c_0$ at step $s_0$, and let $TERM_1$ be the subset of reachable configurations that are terminal.  Define $REACH_2$ to be the union of all reachable configurations from each possible starting configuration attained by adding $s_1$ to a configuration in $TERM_1$.  Let $TERM_2$ be the subset of these reachable configurations that are terminal.
Similarly, define $REACH_i$ to be the union of all reachable sets attained by using initial configuration $c_{i - 1}$ at step $s_{i - 1}$ plus any element of $TERM_{i-1}$, and let $TERM_i$ denote the subset of these configurations that are terminal.

The set of reachable configurations for a $k$-step CRN is the set $REACH_{k}$, and the set of terminal configurations is $TERM_k$. A classical CRN can be represented as a step CRN with $k=1$ steps and an initial configuration of $I=s_0$.

Note that our definitions assume only the terminal configurations of a given step are passed on to seed the subsequent step.  This makes sense if we assume we are dealing with \emph{bounded} systems, as this represents simply waiting long enough for all configurations to reach a terminal state before proceeding to the next step. In this paper we only consider bounded void rule systems; we leave more general definitions to be discussed in future work.

\vspace{-.2cm}
\subsection{Computing Functions in Step CRNs}

Here, we define what it means for a step CRN to compute a function $f(x_1,\ldots_n) = (y_1,\ldots y_m)$ that maps $n$-bit strings to $m$-bit strings.  For each input bit, we denote two separate species types, one representing bit 0, and the other bit 1.  We add one copy for each bit to encode an input $n$-bit strig.  Similarly, each output bit has two representatives(for 0 and 1), and we say the step CRN computes function $f$ if for any given $n$-bit input $x_1,\ldots x_n$, the system results in a final configuration whose output species encode the string $f(x_1,\ldots x_n)$. For a fixed function $f$, the values denoted at each step $s_i$ are fixed to disallow outside computation. 






\para{Input-Strict Step CRN Computing.} Given a Boolean function $f(x_1,\ldots , x_n) = (y_1,\ldots , y_m)$ that maps a string of $n$ bits 
to a string of $m$ bits, 
we define the computation of $f$ with a step CRN. An input-strict step CRN computer is a tuple $C_s = (S,X,Y)$ where $S = ( (\Lambda,\Gamma), (s_0, s_1, \ldots , s_{k-1}))$ is a step CRN, and $X=( (x^0_1,x^1_1),\ldots ,(x^0_n,x^1_n))$ and $Y=( (y^0_1,y^1_1),\ldots , (y^0_m,y^1_m))$ are sequences of ordered-pairs with each $x^0_i,x^1_i,y^0_j,y^1_j \in \Lambda$. Given an $n$-input bit string $b=b_1,\ldots , b_n$, configuration $X(b)$ is defined as the configuration over $\Lambda$ obtained by including one copy of $x^0_i$ only if $b_i=0$ and one copy of $x^1_i$ only if $b_i=1$ for each bit $b_i$. We consider this representation of the input to be strict, as opposed to allowing multiple copies of each input bit species.
The corresponding step CRN $(\Lambda, \Gamma, (s_0 + X(b), \ldots , s_{k-1}))$ is obtained by adding $X(b)$ to $s_0$ in the first step, which conceptually represents the system programmed with specific input $b$.

An input-strict step CRN computer \emph{computes} function $f$ if, for all $n$-bit strings $b$ and for all terminal configurations of $(\Lambda, \Gamma, (s_0 + X(b), \ldots , s_{k-1}))$, the terminal configuration contains at least 1 copy of $y^0_j$ and 0 copies of $y^1_j$ if the $j^{th}$ bit of $f(b)$ is 0, and at least 1 copy of $y^1_j$ and 0 copies of $y^0_j$ if the $j^{th}$ bit of $f(b)$ is 1, for all $j$ from $1$ to $m$.

We use the term \emph{strict} to denote requiring exactly one copy of each bit species and we leave it for future work to consider a more general form of input allowance or strict output. Here, we only consider input-strict computation, so we use input-strict and strict interchangeably.




\para{Relation to CRN Computers.} Previous models of CRN computers considered functions over large domains such as the positive integers.  Due to the infinite domain, the input to such systems cannot be bounded.  As such, the CRN computers shown in \cite{chen2014deterministic} define the input in terms of the volume of some input species.  In these scenarios, CRN computers are limited to computing semi-linear functions.  Here, we instead focus on computing $n$-bit functions, and instead encode the input per bit with potentially unique species. This is a model more similar to the PSPACE computer shown in \cite{thachuk2012space}. 





\vspace{-.2cm}
\subsection{Boolean and Threshold Circuits}
\vspace{-.1cm}

A Boolean circuit on $n$ variables $x_1, x_2, \ldots, x_n$ is a directed, acyclic multi-graph.  The vertices of the graph are generally referred to as \emph{gates}.  The in-degree and out-degree of a gate are called the \emph{fan-in} and \emph{fan-out} of the gate respectively.  The fan-in 0 gates  (\emph{source} gates) are labeled from $x_1, x_2, \ldots, x_n$, or labeled by constants starting at $0$ or $1$.  Each non-source gate is labeled with a function name, such as AND, OR, or NOT.  
Given an assignment of boolean values to variables $x_1, x_2, \ldots, x_n$, each gate in the circuit can be assigned a value by first assigning all source vertices the value matching the labeled constant or labeled variable value and subsequently assigning each gate the value computed by its labeled function on the values of its children.  Given a fixed ordering on the output gates, the sequence of bits assigned to the output gates denotes the value computed by the circuit on the given input.  

The \emph{depth} of a circuit is the longest path from a source vertex to an output vertex. 
Here, we focus on circuits that consist of AND, OR, NOT, and MAJORITY gates with arbitrary fan in. We refer to circuits that use these gates as \textit{threshold circuits} (TC).

\subsection{Notation}  When discussing a boolean circuit, we use the following variables to denote the properties of the circuit: Let $D$ denote the circuit's depth, $G$ the number of gates in the circuit, $W$ the circuit's width, $F_{out}$ the maximum fan-out of any gate in the circuit, $F_{in}$ the maximum fan-in, and $F_{maj}$ the maximum fan-in of any majority gate within the circuit.  
\vspace{-.2cm}
\section{Computation of Threshold Circuits with (3, 0) Rules}\label{sec:3_0_circuits}
\vspace{-.2cm}

Here, we introduce a step CRN system construction with only true void rules that can compute Threshold Circuits. 
In other words, given any TC and some truth assignment to the input variables, we can construct a step CRN with only true void rules that computes the same output as the circuit.

This section specifically focuses on step CRNs consisting of $(3, 0)$ rules. Subsection \ref{3_0_gates} shows how the system can compute individual logic gates, and we show an example construction of a circuit in Subsection \ref{3_0_examples}. We then present the general construction of computing TC circuits by two different methods, differing in the number of species needed based on the fan-out and width of the circuit. 
This results in Theorem \ref{(3,0)_circuit}, which states that TCs can be strictly computed, even with unbounded fan-out, with $O(\min(W^2, G F_{out}))$ species, $O(D \log F_{out})$ steps, and $O(W)$ volume.

\vspace{-.2cm}
\subsection{Computing Logic Gates}\label{3_0_gates}
\vspace{-.2cm}

\para{Indexing.} The number of steps to compute an individual depth level of a circuit varies between 2-8 steps depending on the gates and wiring of the specified circuit. 
To convert a circuit into a $(3,0)$ step CRN system, we need to \emph{index} the wires (input and output) at each level of the circuit in order to ensure the species is input to the correct gate. An example circuit with bit/wire indexing is shown in Figure \ref{fig:(3,0)_example}. At each level, we call the indices of the inputs of gates the \emph{input indices}, and the indices of the output of each gate the \emph{gate indices}. Note that the index numbers may need to change along the wire, or change due to fan-out/fan-in (see Figure \ref{fig:(3,0)_example}). This is accomplished by rules of the form $t_{j\rightarrow i}$ that map an input index of $j$ to a gate index of $i$.

\para{Bit Representation.} The input bits of a binary gate 
are represented in a step CRN with $(3, 0)$ rules by the species $x_n^b$, where $n \in \mathbb{N}$ and $b \in \{ T, F \}$. Here, $n$ represents the bit's index (based on the ordering of all bits into the gates)
and $b$ represents its truth value. Let $f_i^{in}$ be the set of all the indices of input bits fanning into a gate at index $i$ (gate indices). 
Let $f_i^{out}$ be the set of all indices of the output bits fanning out of a gate at index $i$.

The output bits of a gate are represented by the species $y_{n, g}^b$, where $n$ is instead the output bit's index (input index of the next level) and $g$ denotes the gate type $g \in \{B, A, O, N, M \}$ (BUFFER, AND, OR, NOT, and MAJORITY). The BUFFER ($B$) represents a signal wire that changes depth without passing through a gate. For example, the outputs of an AND gate, an OR gate, and a NOT gate at index $n$ are represented by the species $y_{n, A}^b$, $y_{n, O}^b$, and $y_{n, N}^b$, respectively.

\para{AND/OR Gate.}\label{3_0_basic_gates}
The general process to compute an AND gate (an OR gate is similar) is given in Table \ref{tab:(3,0)_AND_OR}. First, all input species are converted into a new species $a_{i,g}^b$ (step 1). The species retains 
truth value $b$ as the original input, and includes the gate index $i$ and type of the gate $g$. The species $b_{i,g}^b$ is then introduced (step 2), which computes the operation of gate $g$ across all existing species. Any species that do not share the same truth value as the gate's intended output are deleted (step 3-4). The species remaining after the operation is then converted into the correct output species (step 5). The species $u_i$, $v_i$, $w_i$, and $t_{j \rightarrow i}$, where $j$ is the input index and $i$ is the gate index, are used to assist in removing excess species in certain steps. 

\begin{figure}[t]
    \vspace{-.2cm}
    \centering
    \includegraphics[width=1.\textwidth]{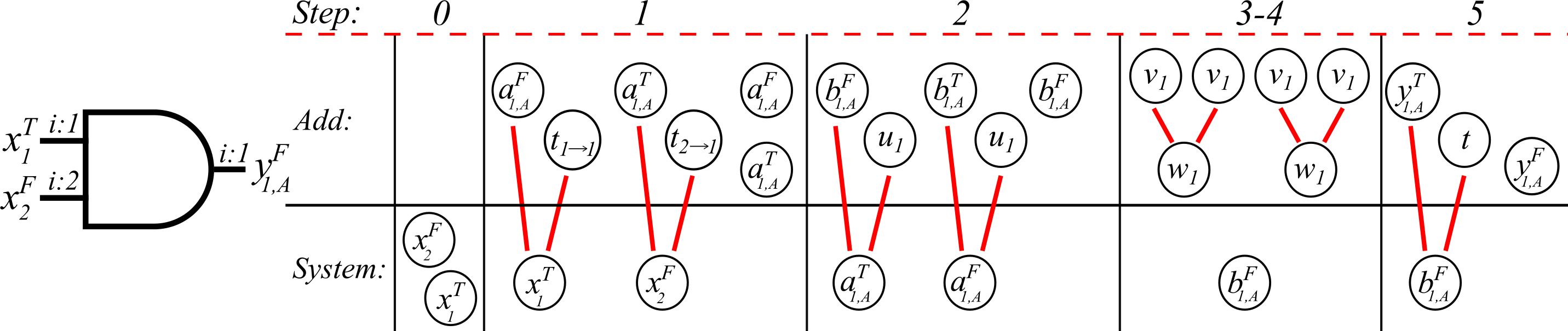}
    \vspace{-.2cm}
    \caption{Example AND gate and steps to compute using $(3, 0)$ rules. Note the gate indexing of the wires ($i:1$ and $i:2$) and the input indexing for the next level ($i:1$ since there is only one gate). The process of computing the gate is shown on the right in steps. The new species added at each step are above and the remaining ones are in the system. The lines show the rules that would be executed during each step. To see the added species and rules in detail, see Table \ref{tab:(3,0)_AND_OR}.}
    \label{fig:andexample}
\end{figure}

\para{AND Example.} Consider an AND gate whose gate index is $1$ with input bits 1 and 0 as shown in Figure \ref{fig:andexample}. In this case, $|f_i^{in}| = 2$ and the initial configuration consists of the species $x_1^T$ and $x_2^F$. Following Table \ref{tab:(3,0)_AND_OR}, this gate can be computed in five steps.
\begin{enumerate}
    \item Two $a_{1,A}^T$, two $a_{1,A}^F$, one $t_{1 \rightarrow 1}$, and one $t_{2 \rightarrow 1}$ species are added to the system. This converts the two input species of the gate into $a_{1,A}^T$ and $a_{1,A}^F$ (causes all species except $a_{1,A}^T$ and $a_{1,A}^F$ to be deleted). 
    \item One $b_{1,A}^T$, two $b_{1,A}^F$, and two $u_1$ species are added. All species except a single $b_{1,A}^F$ are deleted by reactions. 
    \item Four $v_1$ species are added to remove excess species. There are none, so no reactions occur. 
    \item Two $w_1$ are added to delete excess species. Now, only a $b_{1,A}^F$ species remains. 
    \item One $y_{1, A}^T$, one $y_{1, A}^F$ and one $t$ species are added. The $b_{1,A}^F$ species cause the $y_{1, A}^T$ and $t$ species to be deleted. The $y_{1,A}^F$ species is the only species remaining, which represents the intended ``false'' output of the AND gate.
\end{enumerate}

\vspace{-.2cm}
\para{NOT Gate.} Table \ref{tab:(3,0)_NOT} shows the general process to computing NOT gates. To compute a NOT gate, only the output species and species $t$ are added in. In NOT gates specifically, the input species and the output species that share the same truth value $b$ remove each other, leaving the complement of the input species as the remaining and correct output species.

\begin{table}[t]
    \renewcommand{\arraystretch}{.5}
    \begin{center}
    \begin{tabular}{| c | c  c | c | c |}\hline
        
        \multicolumn{3}{|c|}{\textbf{Steps}} & \textbf{Relevant Rules} & \textbf{Description} \\ \hline
        
        \textbf{1} & \textit{Add} & \makecell{\vspace{-8pt} \\ $|f_i^{in}| \cdot a_{i,g}^T$ \\ $|f_i^{in}| \cdot a_{i,g}^F$ \\ $\forall j \in f_i^{in}:$ $t_{j \rightarrow i}$ \\ \vspace{-10pt} } 
        
        & \makecell{$\forall j \in f_i^{in}:$ \\ $x_j^T + a_{i,g}^F + t_{j \rightarrow i} \rightarrow \emptyset$ \\ $x_j^F + a_{i,g}^T + t_{j \rightarrow i} \rightarrow \emptyset$}
        
        & \makecell{$\forall j \in f_i^{in}$, convert $x_{j}^b$ input \\ species into $a_{i,g}^b$ species.} \\ \hline
        
        \textbf{2} & \textit{Add} & \makecell{$b_{i,g}^T$ \\ $|f_i^{in}| \cdot b_{i,g}^F$ \\ $|f_i^{in}| \cdot u_i$} 
        
        & \makecell{$u_i + a_{i,g}^T + b_{i,g}^F \rightarrow \emptyset$ \\ $u_i + a_{i,g}^F + b_{i,g}^T \rightarrow \emptyset$}
        
        & \makecell{Keep at least one of the correct output \\ species and delete all incorrect species. \\ This step computes the AND gate.$^*$ } \\ \hline
        
        \textbf{3} & \textit{Add} & \makecell{\vspace{-8pt} \\ $2|f_i^{in}| \cdot v_i$ \\ \vspace{-8pt} \\} 
        
        & \makecell{$u_i + v_i + v_i \rightarrow \emptyset$}
        
        & \makecell{Delete extra/unwanted species.} \\ \hline
        
        \textbf{4} & \textit{Add} & \makecell{$|f_i^{in}| \cdot w_i$} 
        
        & \makecell{$w_i + v_i + v_i \rightarrow \emptyset$ \\ $w_i + a_{i,g}^F + b_{i,g}^F \rightarrow \emptyset$ \\ \vspace{-9pt}}
        
        & \makecell{Delete extra/unwanted species.} \\ \hline
        
        \textbf{5} & \textit{Add} & \makecell{\vspace{-8pt} \\ $y_{i,g}^T$, $y_{i,g}^F$, $t$ \vspace{-7pt} \\ \\} 
        
        & \makecell{$b_{i,g}^T + y_{i, g}^F + t \rightarrow \emptyset$ \\ $b_{i,g}^F + y_{i,g}^T + t \rightarrow \emptyset$}
        
        & \makecell{Convert $b_{i,g}^b$ into the proper output \\ species $y_{i,g}^b$.} \\ \hline
    
    \end{tabular}
    
    \caption{\textit{(3, 0) rules and steps for an AND gate. To compute an OR gate, add $|f_i^{in}| \cdot b_{i,g}^T$ and one $b_{i,g}^F$ in step 2 instead, and replace $w_i + a_{i,g}^F + b_{i,g}^F \rightarrow \emptyset$ with $w_i + a_{i,g}^T + b_{i,g}^T \rightarrow \emptyset$ in step 4.}}\label{tab:(3,0)_AND_OR}
    \end{center}
    \vspace{-.4cm}
\end{table}

\begin{table}[t]
    \vspace{-.2cm}
    \renewcommand{\arraystretch}{.5}
    \begin{center}
    \begin{tabular}{| c | c  c | c | c |}\hline
    
        \multicolumn{3}{|c|}{\textbf{Steps}} & \textbf{Relevant Rules} & \textbf{Description} \\ \hline
        
        \textbf{1} & \textit{Add} & \makecell{\vspace{-8pt} \\ $y_{i,N}^T$ \\ $y_{i,N}^F$ \\ $t_{j \rightarrow i}$} 
        
        & \makecell{$y_{i, N}^T + x_j^T + t_{j \rightarrow i} \rightarrow \emptyset$ \\ $y_{i, N}^F + x_j^F + t_{j \rightarrow i} \rightarrow \emptyset$} 
        
        & \makecell{The output species ($y_{i,N}^b$) that is \\ the complement of the input species ($x_j^b$) \\ will be the only species remaining.} \\ \hline
    
    \end{tabular}
    \caption{\textit{(3, 0) rules and steps for a NOT gate.}}\label{tab:(3,0)_NOT}
    \end{center}
\end{table}

\para{Majority Gate.}\label{3_0_majority_gates}
The majority gate outputs 1 if and only if more than half of its inputs are 1. Otherwise, it returns 0. The general step process is overviewed in Table \ref{tab:(3,0)_MAJ} 
To compute a majority gate, all input species are converted into a new species $a_{i,M}^b$ (step 1). The species retains the same index $i$ and truth value $b$ as the original input. If the number of species fanning into the majority gate is even, an extra \emph{false} input species is added in. The species $b_{i,M}^b$ is then introduced, which computes the majority operation across all existing species. Any species that represent the minority inputs are deleted (step 2). The species remaining after the operation are converted into the correct output (gate index) species (step 5). The species $u_i$, $v_i$, $w_i$, and $t_{j \rightarrow i}$, where $j$ is the input index and $i$ is the gate index, are used to assist in removing excess species in certain steps.

\begin{table}[t]
\vspace{-.3cm}
\renewcommand{\arraystretch}{.4}
\begin{center}
\begin{tabular}{| c | c  c | c | c |}\hline
    
    \multicolumn{3}{|c|}{\textbf{Steps}} & \textbf{Relevant Rules} & \textbf{Description} \\ \hline
    
    \textbf{1} & \textit{Add} & \makecell{\vspace{-8pt} \\ $|f_i^{in}| \cdot a_{i,M}^T$ \\ $|f_i^{in}| \cdot a_{i,M}^F$ \\ $\forall j \in f_i^{in}:$  $t_{j \rightarrow i}$ \\ \vspace{-10pt} \\} 
    
    & \makecell{$\forall j \in f_i^{in}:$ \\ $x_j^T + a_{i,M}^F + t_{j \rightarrow i} \rightarrow \emptyset$ \\ $x_j^F + a_{i,M}^T + t_{j \rightarrow i} \rightarrow \emptyset$}
    
    & \makecell{$\forall j \in f_i^{in}$, convert $x_{j}^b$ input \\ species into $a_{i,M}^b$ species.} \\ \hline
    
    \textbf{2} & \textit{Add} & \makecell{$\lfloor |f_i^{in}|/2 \rfloor \cdot b_{i,M}^T$ \\ $\lfloor |f_i^{in}|/2 \rfloor \cdot b_{i,M}^F$ \\ $(|f_i^{in}|-1) \cdot u_i$} 
    
    & \makecell{$u_i + a_{i,M}^T + b_{i,M}^F \rightarrow \emptyset$ \\ $u_i + a_{i,M}^F + b_{i,M}^T \rightarrow \emptyset$}
    
    & \makecell{Adding $\lfloor |f_i^{in}|/2 \rfloor$ amounts of $b_{i,M}^T$ and \\ $b_{i,M}^F$ species will delete all of the \\ minority species, leaving some amount \\ of the majority species remaining.} \\ \hline
    
    \textbf{3} & \textit{Add} & \makecell{\vspace{-8pt} \\ $2(|f_i^{in}|-1) \cdot v_i$ \\ \vspace{-7pt} \\} 
    
    & \makecell{$u_i + 2v_i \rightarrow \emptyset$}
    
    & \makecell{Delete extra/unwanted species.} \\ \hline
    
    \textbf{4} & \textit{Add} & \makecell{$(|f_i^{in}|-1) \cdot w_i$} 
    
    & \makecell{$w_i + 2v_i \rightarrow \emptyset$ \\ $w_i + a_{i,M}^T + b_{i,M}^T \rightarrow \emptyset$ \\ $w_i + a_{i,M}^F + b_{i,M}^F \rightarrow \emptyset$ \\ \vspace{-9pt}}
    
    & \makecell{Delete extra/unwanted species.} \\ \hline
    
    \textbf{5} & \textit{Add} & \makecell{\vspace{-8pt} \\ $y_{i,M}^T$ \\ $y_{i,M}^F$ \\ $t$} 
    
    & \makecell{$a_{i,M}^T + y_{i,M}^F + t \rightarrow \emptyset$ \\ $a_{i,M}^F + y_{i,M}^T + t \rightarrow \emptyset$}
    
    & \makecell{Convert $a_{i,M}^b$ into the proper output \\ species ($y_{i,M}^b$).} \\ \hline
    
\end{tabular}
\caption{\textit{(3, 0) rules and steps for a majority gate.}}\label{tab:(3,0)_MAJ}
\end{center}
\vspace{-1.cm}
\end{table}

\vspace{-.2cm}

\subsection{(3,0) Circuit Example}\label{3_0_examples}
\vspace{-.2cm}

\begin{table}[t]
\vspace{-.2cm}
\centering\renewcommand{\arraystretch}{1.1}
\hspace{-.9cm}
\begin{tabular}{|c|l|l|@{}l@{}|c|l|l|} 
\hline
\multicolumn{3}{|r}{\multirow{2}{*}{\textbf{Initial Configuration:}}} & \multicolumn{1}{c}{} & \multicolumn{3}{l|}{\multirow{2}{*}{$y_{1,B}^T$ \hspace{4pt} $y_{2,B}^T$ \hspace{4pt} $y_{3,B}^T$ \hspace{4pt} $y_{4,B}^F$}} \\
\multicolumn{3}{|r}{} & \multicolumn{1}{c}{} & \multicolumn{3}{l|}{} \\ 
\hline
\multicolumn{2}{|c|}{\textbf{Steps}} & \multicolumn{1}{c|}{\textbf{Relevant Rules}} &  & \multicolumn{2}{c|}{\textbf{Steps}} & \multicolumn{1}{c|}{\textbf{Relevant Rules}} \\ 
\cline{1-3}\cline{5-7}
    \multirow{4}{*}{\textbf{1}} 
    & \multirow{4}{*}{\begin{tabular}[c]{@{}l@{}}$x_1^T$, $x_2^T$, $x_3^T$, $x_4^T$\\ $t_{1 \rightarrow 1}$, $t_{3 \rightarrow 3}$\\ $x_1^F$, $x_2^F$, $x_3^F$, $x_4^F$\\ $t_{2 \rightarrow 2}$, $t_{4 \rightarrow 4}$\end{tabular}} 
    & \multirow{4}{*}{\begin{tabular}[c]{@{}l@{}}$y_{1,B}^T + x_1^F + t_{1 \rightarrow 1} \rightarrow \emptyset$\\ $y_{2,B}^T + x_2^F + t_{2 \rightarrow 2} \rightarrow \emptyset$\\ $y_{3,B}^T + x_3^F + t_{3 \rightarrow 3} \rightarrow \emptyset$\\ $y_{4,B}^F + x_4^T + t_{4 \rightarrow 4} \rightarrow \emptyset$\end{tabular}} &  & 
    \multirow{4}{*}{\textbf{10}} & \multirow{4}{*}{\begin{tabular}[c]{@{}l@{}}$2 a_{1,A}^T$, $2 a_{2,A}^T$\\ $2 a_{1,A}^F$, $2 a_{2,A}^F$\\ $t_{2 \rightarrow 1}$, $t_{4 \rightarrow 2}$\\$t_{1 \rightarrow 1}$, $t_{3 \rightarrow 2}$\end{tabular}} & \multirow{4}{*}{\begin{tabular}[c]{@{}l@{}}$x_1^F + a_{1,A}^T + t_{1 \rightarrow 1} \rightarrow \emptyset$\\ $x_2^T + a_{1,A}^F + t_{2 \rightarrow 1} \rightarrow \emptyset$\\ $x_3^T + a_{2,A}^F + t_{3 \rightarrow 2} \rightarrow \emptyset$\\ $x_4^F + a_{2,A}^T + t_{4 \rightarrow 2} \rightarrow \emptyset$\end{tabular}}\\
 &  &  & \multicolumn{1}{c|}{\textbf{}} &  &  &  \\
 &  &  & \multicolumn{1}{c|}{\textbf{}} &  &  &  \\
 &  &  & \multicolumn{1}{c|}{\textbf{}} &  &  &  \\
\cline{1-3}\cline{5-7}
\multirow{4}{*}{\textbf{2}} 
    & \multirow{4}{*}{\begin{tabular}[c]{@{}l@{}}$y_{1,N}^T$, $2 a_{2,O}^T$, $y_{3,B}^T$\\ $t_{1 \rightarrow 1}$, $t_{3 \rightarrow 2}$\\ $y_{1,N}^F$, $2 a_{2,O}^F$, $y_{3,B}^F$\\ $t_{2 \rightarrow 2}$, $t_{4 \rightarrow 3}$\end{tabular}} 
    & \multirow{4}{*}{\begin{tabular}[c]{@{}l@{}}$x_1^T + y_{1,N}^T + t_{1 \rightarrow 1} \rightarrow \emptyset$\\ $x_2^T + a_{2,O}^F + t_{2 \rightarrow 2} \rightarrow \emptyset$\\ $x_3^T + a_{2,O}^F + t_{3 \rightarrow 2} \rightarrow \emptyset$\\ $x_4^T + y_{3,B}^T + t_{4 \rightarrow 3} \rightarrow \emptyset$\end{tabular}} 
    &\multicolumn{1}{c|}{\textbf{}} 
    & \multirow{4}{*}{\textbf{11}} & \multirow{4}{*}{\begin{tabular}[c]{@{}l@{}}$b_{1,A}^T$, $b_{2,A}^T$\\ $2 u_1$, $2 b_{1,A}^F$\\ $2 b_{2,A}^F$, $2 u_2$\end{tabular}} & \multirow{4}{*}{\begin{tabular}[c]{@{}l@{}}$a_{1,A}^T + b_{1,A}^F + u_1 \rightarrow \emptyset$\\ $a_{1,A}^F + b_{1,A}^T + u_1 \rightarrow \emptyset$\\ $a_{2,A}^T + b_{2,A}^F + u_2 \rightarrow \emptyset$\\ $a_{2,A}^F + b_{2,A}^T + u_2 \rightarrow \emptyset$\end{tabular}} \\
     &  &  & \multicolumn{1}{c|}{\textbf{}} &  &  &  \\
 &  &  & \multicolumn{1}{c|}{\textbf{}} &  &  &  \\
 &  &  & \multicolumn{1}{c|}{\textbf{}} &  &  &  \\
    \cline{1-3}\cline{5-7}
   \multirow{1}{*}{\textbf{3}} & \multirow{1}{*}{\begin{tabular}[c]{@{}l@{}}$2 b_{2,O}^T$, $2 u_2$, $b_{2,O}^F$\end{tabular}} & \multirow{1}{*}{$a_{2,O}^T + b_{2,O}^F + u_2 \rightarrow \emptyset$}  & &  \multicolumn{1}{l|}{\multirow{1}{*}{\textbf{12}}} 
    & \multirow{1}{*}{$4 v_1$, $4 v_2$} & \multirow{1}{*}{\textit{No Rules Apply}} \\
    \cline{1-3}\cline{5-7}
    
\multirow{1}{*}{\textbf{4}} & \multirow{1}{*}{$4 v_2$} & \multirow{1}{*}{$u_2 + v_2 + v_2 \rightarrow \emptyset$}   &    & \multirow{2}{*}{\textbf{13}} & \multirow{2}{*}{$2 w_1$, $2 w_2$} & \multirow{2}{*}{\begin{tabular}[c]{@{}l@{}}$w_1 + v_1 + v_1 \rightarrow \emptyset$\\ $w_2 + v_2 + v_2 \rightarrow \emptyset$\end{tabular}} \\
 \cline{1-3}
 \multirow{2}{*}{\textbf{5}} & \multirow{2}{*}{$2 w_2$} & \multirow{2}{*}{\begin{tabular}[c]{@{}l@{}}$w_2 + v_2 + v_2 \rightarrow \emptyset$\\ $w_2 + a_{2,O}^T + b_{2,O}^T \rightarrow \emptyset$\end{tabular}}  &  &  &  &  \\
\cline{5-7}
 &  &  &    & \multicolumn{1}{l|}{\multirow{2}{*}{\textbf{14}}} & \multirow{2}{*}{\begin{tabular}[c]{@{}l@{}}$y_{1,A}^T$, $y_{2,A}^T$, $2 t$\\ $y_{1,A}^F$, $y_{2,A}^F$\end{tabular}} & \multirow{2}{*}{\begin{tabular}[c]{@{}l@{}}$b_{1,A}^F + y_{1,A}^T + t \rightarrow \emptyset$\\ $b_{2,A}^F + y_{2,A}^T + t \rightarrow \emptyset$\end{tabular}} \\
 \cline{1-3}
\multirow{1}{*}{\textbf{6}} & \multirow{1}{*}{\begin{tabular}[c]{@{}l@{}}$y_{2,O}^T$, $t$, $y_{2,O}^F$\end{tabular}} & \multirow{1}{*}{$b_{2,O}^T + y_{2,O}^F + t \rightarrow \emptyset$} & & & & \\
\cline{1-3}\cline{5-7}
 \multirow{2}{*}{\textbf{7}} & \multirow{2}{*}{\begin{tabular}[c]{@{}l@{}}$y_{2,O}^T$, $r$, $y_{2,O}^F$\end{tabular}} & \multirow{2}{*}{$y_{2,O}^T + y_{2,O}^T + r \rightarrow \emptyset$} &  & \multirow{2}{*}{\textbf{15}} & \multirow{2}{*}{\begin{tabular}[c]{@{}l@{}}$x_1^T$, $x_2^T$, $t_{1 \rightarrow 1}$\\ $x_1^F$, $x_2^F$, $t_{2 \rightarrow 2}$\end{tabular}} & \multirow{2}{*}{\begin{tabular}[c]{@{}l@{}}$y_{1,A}^F + x_1^T + t_{1 \rightarrow 1} \rightarrow \emptyset$\\ $y_{2,A}^F + x_2^T + t_{2 \rightarrow 2} \rightarrow \emptyset$\end{tabular}} \\
  & & & & & &\\
  \cline{1-3}\cline{5-7}
  \multirow{2}{*}{\textbf{8}} & \multirow{2}{*}{$2 y_{2,O}^T$, $2 y_{2,O}^F$} & \multirow{2}{*}{$y_{2,O}^F + y_{2,O}^F + y_{2,O}^F \rightarrow \emptyset$} & &\multirow{2}{*}{\textbf{16}} & \multirow{2}{*}{\begin{tabular}[c]{@{}l@{}}$2 a_{1,O}^T$, $t_{1 \rightarrow 1}$ \\$2 a_{1,O}^F$, $t_{2 \rightarrow 1}$\end{tabular}} & \multirow{2}{*}{\begin{tabular}[c]{@{}l@{}}$x_1^F + a_{1,O}^T + t_{1 \rightarrow 1} \rightarrow \emptyset$\\ $x_2^F + a_{1,O}^T + t_{2 \rightarrow 1} \rightarrow \emptyset$\end{tabular}} \\
  & & & & & &\\
 \cline{1-3}\cline{5-7}
\multirow{5}{*}{\textbf{9}} & \multirow{5}{*}{\begin{tabular}[c]{@{}l@{}}$x_1^T$, $x_2^T$, $x_3^T$, $x_4^T$\\ $t_{1 \rightarrow 1}$, $t_{2 \rightarrow 3}$\\ $x_1^F$, $x_2^F$, $x_3^F$, $x_4^F$\\ $t_{2 \rightarrow 2}$, $t_{3 \rightarrow 4}$\end{tabular}} & \multirow{5}{*}{\begin{tabular}[c]{@{}l@{}}$y_{1,N}^F + x_1^T + t_{1 \rightarrow 1} \rightarrow \emptyset$\\ $y_{2,O}^T + x_2^F + t_{2 \rightarrow 2} \rightarrow \emptyset$\\ $y_{2,O}^T + x_3^F + t_{2 \rightarrow 3} \rightarrow \emptyset$\\ $y_{3,B}^F + x_4^T + t_{3 \rightarrow 4} \rightarrow \emptyset$\end{tabular}} & & \multicolumn{1}{l|}{\multirow{1}{*}{\textbf{17}}} & \multirow{1}{*}{\begin{tabular}[c]{@{}l@{}}$2 b_{1,O}^T$, $2 u_1$, $b_{1,O}^F$\end{tabular}} & \multirow{1}{*}{$a_{1,O}^F + b_{1,O}^T + u_1 \rightarrow \emptyset$}  \\
   \cline{5-7}
   &  &  &  &  \multirow{1}{*}{\textbf{18}} & \multirow{1}{*}{$4 v_1$} & \multirow{1}{*}{\textit{No Rules Apply}}  \\ 
\cline{5-7} 
 &  &  &  & \multirow{1}{*}{\textbf{19}} & \multirow{1}{*}{$2 w_1$} & \multirow{1}{*}{$w_1 + v_1 + v_1 \rightarrow \emptyset$} \\ 
\cline{5-7}
 &  &  &  & \multicolumn{1}{l|}{\multirow{1}{*}{\textbf{20}}} & \multirow{1}{*}{\begin{tabular}[c]{@{}l@{}}$y_{1,O}^T$, $t$, $y_{1,O}^F$\end{tabular}} & \multirow{1}{*}{$b_{1,O}^F + y_{1,O}^T + t \rightarrow \emptyset$} \\ 
\cline{5-7}
 &  &  &  & \multirow{1}{*}{\textbf{21}} & \multirow{1}{*}{\begin{tabular}[c]{@{}l@{}}$x_1^T$, $t_{1 \rightarrow 1}$ $x_1^F$\end{tabular}} & \multirow{1}{*}{$y_{1,O}^F + x_1^T + t_{1 \rightarrow 1} \rightarrow \emptyset$} \\

\hline
\end{tabular}

\caption{$(3, 0)$ rules and steps to compute the circuit in Figure \ref{fig:(3,0)_example} based on the indexing shown in Figure \ref{fig:(3,0)_figure}a. Note that as in other tables, the `Steps' column shows the number and types of species being added at the beginning of that step.}
\label{tab:(3,0)_example}
\end{table}

With the computation of individual gates demonstrated in our system, we now expand these features to computing entire circuits. We begin with a simple example (Figure \ref{fig:(3,0)_example}) to show the concepts before giving the general construction.
The circuit has four inputs: $x_1$, $x_2$, $x_3$, and $x_4$. At the first depth layer, $x_1$ fans into a NOT gate and $x_2$ and $x_3$ are both fanned into an OR gate. At the next depth level, the output of the OR gate is fanned out twice. One of these outputs, along with the output of the NOT gate, is fanned into an AND gate, while the other and $x_4$ fans into another AND gate. At the last depth level, both AND gate outputs fan into an OR gate, which computes the final output of the circuit.

Table \ref{tab:(3,0)_example} shows how to compute the circuit in Figure \ref{fig:(3,0)_example}. The primary inputs of the circuit in Figure \ref{fig:(3,0)_example} are represented by the species in the initial configuration. Step 1 converts the primary inputs into input species. If there was any fan out of the primary inputs, it would be done in this step. Steps 2-6 compute the gates at the first depth level. Steps 7-8 compute the fan out between the first and second depth level. Step 9 converts the outputs of the gates at the first depth level into input species. Steps 10-14 use those inputs to compute the gates at the second depth level. Step 15 converts the outputs of these gates into inputs. Steps 16-20 compute the final gate. Step 21 converts the output of that gate into an input species that represents the solution to the circuit ($x_1^F$).

\vspace{-.2cm}
\subsection{Computing Circuits}\label{3_0_computation}
\vspace{-.2cm}

\begin{figure}[t]
    \vspace{-.2cm}
    \centering
    \begin{subfigure}[b]{0.22\textwidth}
         \centering
         \includegraphics[width=1.\textwidth]{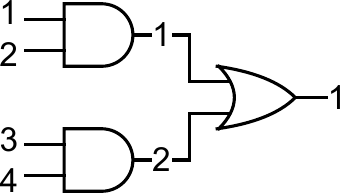}
         \caption{}\label{fig:aoc2}
    \end{subfigure}
    \begin{subfigure}[b]{0.22\textwidth}
         \centering
         \includegraphics[width=1.\textwidth]{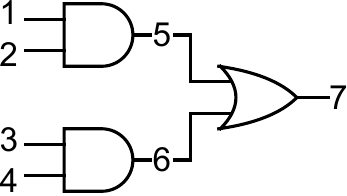}
         \caption{}\label{fig:aoc1}
    \end{subfigure}
    \begin{subfigure}[b]{0.25\textwidth}
         \centering
         \includegraphics[width=1.\textwidth]{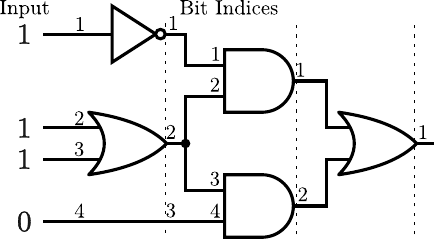}
         \caption{}\label{fig:(3,0)_example}
    \end{subfigure}
    \begin{subfigure}[b]{0.22\textwidth}
        \centering
        \includegraphics[width=.9\textwidth]{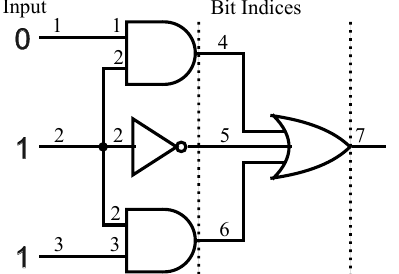}
        \caption{}\label{fig:(2,0)_example}
    \end{subfigure}
    \vspace{-.2cm}
    \caption{(a) Example indexing pattern of wires for the step CRN method using $O(W^2)$ species. (b) Example indexing pattern of wires for the step CRN method using $O(G)$ species. (c) Example circuit (with indexing) for Table \ref{tab:(3,0)_example}. (d) Example circuit (with indexing) for Table \ref{tab:(2,0)_example}.}\label{fig:(3,0)_figure}
    \vspace{-.2cm}
\end{figure}

\begin{lemma}\label{(3,0)_W2_circuit}
Threshold circuits (TC) with a max fan-out of 2 can be strictly computed by a step CRN with only (3,0) rules, $O(W^2)$ species, $O(D)$ steps, and $O(W)$ volume.
\end{lemma}
\begin{proof}
The initial configuration of the step CRN should consist of one $y_{n,B}^b$ species for each primary input with the appropriate indices and truth values. Section \ref{3_0_gates} explains how to compute TC gates. In order to run a circuit, we need to convert the outputs at an index $i$ into the inputs for the next gate at index $j$. To simulate circuits with $O(W^2)$ species, we also need to be able to reuse these input, output, and helper species. This can be accomplished by having unique species for each gate at a given depth level. Figure \ref{fig:aoc2} shows a pattern the gates can be indexed in. 

When reusing species, we incorporate a unique $t_{i \rightarrow j}$ species (different from the $t_{j \rightarrow i}$ species used in computing gates) for each gate at index $i$ that converts the output species into an input species with the index $j$. Converting outputs into inputs is done for all gates at the same depth level. Table \ref{tab:(3,0)_convert} shows the steps and rules needed to complete this process.

\begin{table}[t]
\renewcommand{\arraystretch}{.5}
\begin{center}
\begin{tabular}{| c | c  c | c | c |}\hline
    
    \multicolumn{3}{|c|}{\textbf{Steps}} & \textbf{Relevant Rules} & \textbf{Description} \\ \hline
    
    \textbf{1} & \textit{Add} & \makecell 
    {\vspace{-10pt} \\ $\forall j \in f_i^{out}:$  $x_j^T, x_j^F, t_{i \rightarrow j}$ \\ \vspace{-10pt}} 
    
    & \makecell{$\forall j \in f_i^{out}:$ \\ $y_{i,g}^T + x_j^F + t_{i \rightarrow j} \rightarrow \emptyset$ \\ $y_{i,g}^F + x_j^T + t_{i \rightarrow j} \rightarrow \emptyset$}
    
    & \makecell{$\forall j \in f_i^{out}$, convert $y_{i,g}^b$ output \\ species into $x_j^b$ input species.} \\ \hline
\end{tabular}
\caption{\textit{(3, 0) rules for converting outputs into inputs per circuit level.}}\label{tab:(3,0)_convert}
\end{center}
\end{table}

\para{Fan Out.} In order to perform a 2-fan out, we create a second copy of the output species that is fanning out. Table \ref{tab:(3,0)_fan_out} shows the steps and rules needed to perform this duplication. After duplicating the output, the simulation continues as usual. All outputs at the same depth level can be fanned out at the same time using these two steps.

\begin{table}[t]
\vspace{-.2cm}
\renewcommand{\arraystretch}{.5}
\begin{center}
\begin{tabular}{| c | c c | c | c |}\hline
    
    \multicolumn{3}{|c|}{\textbf{Steps}} & \textbf{Relevant Rules} & \textbf{Description} \\ \hline
    
    \textbf{1} & \textit{Add} & \makecell{\vspace{-7pt} \\ $y_{i,g}^T, y_{i,g}^F, r$} & \makecell{$y_{i,g}^T + y_{i,g}^T + r \rightarrow \emptyset$ \\ $y_{i,g}^F + y_{i,g}^F + r \rightarrow \emptyset$} & \makecell{Flip output's bit (e.g. if species \\ $y_{i,g}^T$ is present, then delete \\ it and preserve $y_{i,g}^F$)} \\ \hline
    
    \textbf{2} & \textit{Add} & \makecell{$2y_{i,g}^T, 2y_{i,g}^F$} & \makecell{$y_{i,g}^T + y_{i,g}^T + y_{i,g}^T \rightarrow \emptyset$ \\ $y_{i,g}^F + y_{i,g}^F + y_{i,g}^F \rightarrow \emptyset$} & \makecell{Delete all copies of the negation \\ of the initial input, and preserve \\ the two copies of the input \\ that were just added.} \\ \hline
    
\end{tabular}
\caption{\textit{(3,0) rules and steps for 2-fan out.}}\label{tab:(3,0)_fan_out}
\end{center}
\end{table}

\para{Complexity.} The $t_{i \rightarrow j}$ approach results in, at most, $W^2$ unique species since $1 \leq i,j \leq W$. All other types of species either have $O(1)$ or $O(W)$ unique species. Therefore, a simulation of a circuit with a max fan-out of 2 requires $O(W^2)$ species.

All gates at a given depth level are evaluated at the same time, so a simulation of a circuit with a max fan-out of 2  requires $O(D)$ steps.
Additionally, circuits are evaluated one depth level at a time. Thus, at most, a max width amount of input, output, and helper species are added at the same time. All of the input, output, and helper species from previous depth levels get deleted when progressing to the next depth level, so the simulation requires $O(W)$ volume.
A constant number of species, steps, and volume are needed to perform a 2-fan out, so a 2-fan out operation does not affect the complexity.
\end{proof}

\begin{lemma}\label{(3,0)_G_circuit}
Threshold circuits (TC) with a max fan-out of 2 can be strictly computed by a step CRN with only $(3, 0)$ rules, $O(G)$ species, $O(D)$ steps, and $O(W)$ volume.
\end{lemma}
\begin{proof}
Most of the rules, species, and steps used to compute a circuit with $O(W^2)$ species (Lemma \ref{(3,0)_W2_circuit}) should also be used for this step CRN. The main difference is that there is an index for every gate in the circuit instead of limiting the indexes of these species by the max width. Figure \ref{fig:aoc1} shows a pattern the gates can be indexed in. Also, we don't need the $t_{i \rightarrow j}$ species. This is because rules could overlap when reusing species, so the $t_{i \rightarrow j}$ species was used to make certain rules distinct and prevent the wrong reactions from occurring. However, each gate being represented by unique species eliminates the possibility of this error as every rule used to compute a gate will also be unique. So, in this step CRN, all instances where $t_{i \rightarrow j}$ species is used (including those in Section \ref{3_0_gates}) are replaced by the generic $t$ species.

\textbf{Complexity.} The $t_{i \rightarrow j}$ species that was the bottleneck for species in the step CRN discussed in Lemma \ref{(3,0)_W2_circuit} has been replaced by the $t$ species. Therefore, the number of species is no longer upper bounded by $O(W^2)$. Instead, there are unique species for each gate, thus requiring $O(G)$ species.
The differences discussed in this lemma do not affect the step or volume complexity determined in Lemma \ref{(3,0)_W2_circuit} ($O(D)$ and $O(W)$, respectively).
\end{proof}

\begin{theorem}\label{(3,0)_circuit}
    Threshold circuits (TC) can be strictly computed by a step CRN with only $(3, 0)$ rules, $O(\min(W^2, G \cdot F_{out}))$ species, $O(D \log F_{out})$ steps, and $O(W)$ volume.
\end{theorem}
\begin{proof}
Since the methods used in Lemmas \ref{(3,0)_W2_circuit} and \ref{(3,0)_G_circuit} can only achieve a max fan-out of 2, a circuit with a larger fan-out ($F_{out}$) must be turned into a circuit with a max fan-out of 2. 
The max width does not change in the transformation process, but the total number of gates does. The transformation process creates a binary tree-like structure within the circuit, where the gate that is fanning out can be likened to the root, and the gates that the output is being inputted into can be likened to the leaves. Therefore, because a binary tree can have, at most, $2\ell -1$ vertices, and an arbitrary fan out already has the ``root'' gate and ``leaves'' gates, the transformation process requires, at most, $\ell -2$ more gates to construct the binary tree-like structure. Thus, the method requiring $O(G)$ species would require $O(GF_{out})$ species to simulate a circuit with arbitrary fan-out. The most efficient method can be used for a given circuit, resulting in $O(\min(W^2, GF_{out}))$ species required to simulate a circuit.

Due to the binary tree-like structure made for gates with large fan out, the transformation process would increase the size of the depth from $D$ to $D \log F_{out}$. Since the steps are dependent on the size of the depth, a circuit simulation would require $O(D \log F_{out})$ steps.

The volume of a circuit simulation does not differ between the two methods ($O(W)$ volume) nor is it affected by the transformation process.
\end{proof}

\vspace{-.4cm}

\section{Threshold Circuits with (2, 0) and (2, 1) Catalyst Rules}\label{sec:2_0_formulas}
\vspace{-.2cm}



Having established computation results with step CRNs using only true void rules, we now examine step CRNs whose rule-sets contain catalytic rules. 
These rulesets can either consist of only $(2,1)$ rules or both $(2,1)$ and $(2,0)$ rules.
Subsection \ref{2_0_gates} shows how the computation of logic gates is possible in step CRNs with just $(2, 0)$ or $(2, 1)$ rules. We then demonstrate with Theorem \ref{2_0_catalysts} how the system can compute TCs with $O(G)$ species, $O(D)$ steps, and $O(W)$ volume.
Subsection \ref{2_1_catalysts} then shows that TCs can also be calculated (with more steps) with only the $(2,1)$ catalyst rules.

\vspace{-.3cm}

\subsection{Computing Logic Gates}\label{2_0_gates}
\vspace{-.2cm}

\para{Bit Representation and Indexing.} The inputs of a binary gate are constructed the same as in Section \ref{3_0_gates}. However, with catalysts, we slightly modify our indexing scheme. When fanning out, we do not split the output of the gate into input species with different indices because the catalyst rules remove the need to differentiate the input species. Let $f_i^{in}$ be the set of all the indices of the inputs fanning into a gate at index $i$. Let $f_i^{out}$ be the set of all the indices of the inputs fanning out of a gate at index $i$.

The output of a gate is represented by the species $y_i^b$ or $y_{j \rightarrow i}^b$, where $j$ is the index of the input bit and $i$ is the index of the gate.


\para{AND/OR/NOT Gate.}\label{2_0_basic_gates} Table \ref{tab:(2,0)_AND_OR_NOT} shows the general process to computing AND, OR, and NOT gates.
To compute an AND gate, we add a single copy of the species representing a true output ($y_{i}^T$) and a species representing a false output for each input  ($\forall j \in f_i^{in}:$ $y_{j \rightarrow i}^F$). To compute an OR gate instead, we add a species representing a true output ($y_{j \rightarrow i}^T$) for each input and a single $y_{i}^F$ species. To compute NOT gates, we add one copy of each output species ($y_i^b$). For every input into an AND/OR/NOT gate, a corresponding rule should be created to remove the output species of the gate with the opposite truth value to the input. If the output species has a unique $j \rightarrow i$ index, then only the input with the corresponding $i$ can delete that output species.

These gates can also be computed with (2,1) catalyst rules by making the $x_j^b$ species a catalyst.  For example, the (2,0) rule $x_j^T + y_i^T \rightarrow \emptyset$ would be replaced by the (2,1) rule $x_j^T + y_i^T \rightarrow x_j^T$.

\para{OR Example.} Consider OR gate whose gate index is 1 with input bits 0 and 1. Here, $|f_i^{in}| = 2$, and the initial configuration consists of the species $x_1^F$ and a $x_2^T$. 

This gate can be computed in one step, following Table \ref{tab:(2,0)_AND_OR_NOT}, by adding one $y_{1 \rightarrow 1}^T$, one $y_{2 \rightarrow 1}^T$, and one $y_{1}^F$ species to the system. The species $x_2^T$ and $y_{1}^F$ delete each other. $x_1^F$ and $y_{1 \rightarrow 1}^T$ are also removed together. Only the species $y_{2 \rightarrow 1}^T$ remains, which represents the intended ``true'' output of the OR gate.


\begin{table}[t]
\vspace{-.2cm}
\renewcommand{\arraystretch}{.5}
\begin{center}
\begin{tabular}{ | c | c c | c | c | c c | c |}\hline

\textbf{Gate Type} & \multicolumn{2}{|c|}{\textbf{Step}} & \textbf{Relevant Rules} & \textbf{Description} \\ \hline

\textbf{AND} & \textit{Add} & \makecell{$y_{i}^T$ \\ $\forall j \in f_i^{in}:$ $y_{j \rightarrow i}^F$} & \makecell{$x_j^T + y_{j \rightarrow i}^F \rightarrow \emptyset$ \\ $x_j^F + y_i^T \rightarrow \emptyset$} & \makecell{An input species with a certain \\ truth value deletes the \\ complement output species.} \\ \hline

 \textbf{OR} & \textit{Add} & \makecell{$y_{i}^F$ \\ $\forall j \in f_i^{in}:$ $y_{j \rightarrow i}^T$} & \makecell{$x_j^T + y_i^F \rightarrow \emptyset$ \\ $x_j^F + y_{j \rightarrow i}^T \rightarrow \emptyset$} & \makecell{An input species with a certain \\ truth value deletes the \\ complement output species.} \\ \hline

 \textbf{NOT} & \textit{Add} & \makecell{$y_{i}^T$ \\ $y_{i}^F$} & \makecell{$x_j^T + y_i^T \rightarrow \emptyset$ \\ $x_j^F + y_i^F \rightarrow \emptyset$} & \makecell{The input and output species that \\ share the same truth value delete \\ each other.} \\ \hline

\end{tabular}

\caption{\textit{(2, 0) rules for AND, OR, and NOT gates.}}\label{tab:(2,0)_AND_OR_NOT}

\end{center}
\end{table}




\vspace{-.2cm}
\para{Majority Gate.}\label{2_0_majority_gates} The general process of computing a majority gate is shown at Table \ref{tab:(2,0)_MAJ}. To compute a majority gate, all input species are converted into a new species $a_{i}^b$ (step 1). The species retains the same truth value $b$ as the original input and has the gate index $i$. If the number of species fanning into the majority gate is even, an extra \emph{false} input species is added in. The species $b_{i}^b$ is then introduced, which computes the majority operation across all existing species. Any species that represent the minority inputs are deleted (step 2). The species remaining afterwards are then converted into the correct output species (step 3).

\begin{table}[t]
\vspace{-.2cm}
\renewcommand{\arraystretch}{.5}
\begin{center}
\begin{tabular}{| c | c  c | c | c |}\hline
    \multicolumn{3}{|c|}{\textbf{Steps}} & \textbf{Relevant Rules} & \textbf{Description} \\ \hline
    
    \textbf{1} & \textit{Add} & \makecell{$|f_i^{in}| \cdot a_i^T$ \\ $|f_i^{in}| \cdot a_i^F$} 
    
    & \makecell{\vspace{-10pt} \\ $\forall j \in f_i^{in}:$ \\ $x_j^T + a_i^F \rightarrow \emptyset$ \\ $x_j^F + a_i^T \rightarrow \emptyset$ \\ \vspace{-9pt}}
    
    & \makecell{$\forall j \in f_i^{in}$, convert $x_{j}^b$ input \\ species into $a_i^b$ species.} \\ \hline
    
    \textbf{2} & \textit{Add} & \makecell{$\lfloor |f_i^{in}|/2 \rfloor \cdot b_i^T$ \\ $\lfloor |f_i^{in}|/2 \rfloor \cdot b_i^F$} 
    
    & \makecell{$a_i^T + b_i^F \rightarrow \emptyset$ \\ $a_i^F + b_i^T \rightarrow \emptyset$}
    
    & \makecell{Adding $\lfloor |f_i^{in}|/2 \rfloor$ amounts of $b_i^T$ and \\ $b_i^F$ species will delete all of the \\ minority species, leaving some amount \\ of the majority species remaining.} \\ \hline
    
    \textbf{3} & \textit{Add} & \makecell{\vspace{-8pt} \\ $y_i^T$ \\ $y_i^F$ \\ \vspace{-10pt}} 
    
    & \makecell{$a_i^T + y_i^F \rightarrow \emptyset$ \\ $a_i^F + y_i^T \rightarrow \emptyset$}
    
    & \makecell{Convert $a_i^b$ into the proper output \\ species ($y_i^b$).} \\ \hline
    
\end{tabular}

\caption{\textit{(2, 0) rules for majority gates.}}\label{tab:(2,0)_MAJ}
\end{center}
\end{table}

\vspace{-.3cm}
\subsection{Examples}\label{2_0_examples}
\vspace{-.2cm}

\begin{table}[t]
\renewcommand{\arraystretch}{1.}  
\begin{center}
    \begin{tabular}{|lllllllll|}
    \hline
    \multicolumn{4}{|r}{\multirow{2}{*}{\textbf{Initial Configuration:}}}  &  & \multicolumn{4}{l|}{\multirow{2}{*}{$y_1^F$  \hspace{4pt}  $y_2^T$  \hspace{4pt}  $y_3^T$}}  \\
    \multicolumn{4}{|r}{}  & \multicolumn{1}{c}{}  & \multicolumn{4}{l|}{}  \\ \hline
    \multicolumn{3}{|c|}{\textbf{Steps}}  & \multicolumn{1}{c|}{\textbf{Relevant Rules}}  & \multicolumn{1}{c|}{} & \multicolumn{3}{c|}{\textbf{Steps}}  & \multicolumn{1}{c|}{\textbf{Relevant Rules}}  \\ \cline{1-4} \cline{6-9} 
    \multicolumn{1}{|l|}{\textbf{1}}  & \textit{Add}  & \multicolumn{1}{l|}{$d_x$}  & \multicolumn{1}{l|}{\textit{No Rules Apply}}  & \multicolumn{1}{l|}{} & \multicolumn{1}{c|}{\textbf{8}}  & \multicolumn{1}{c}{\textit{Add}}  & \multicolumn{1}{l|}{$d_x$}  & $d_x + d_x \rightarrow \emptyset$  \\ \cline{1-4} \cline{6-9} 
    \multicolumn{1}{|l|}{\textbf{2}}  & \textit{Add}  & \multicolumn{1}{l|}{$d_x$}  & \multicolumn{1}{l|}{$d_x + d_x \rightarrow \emptyset$}  & \multicolumn{1}{l|}{} & \multicolumn{1}{c|}{\multirow{3}{*}{\textbf{9}}}  & \multicolumn{1}{c}{\multirow{3}{*}{\textit{Add}}} & \multicolumn{1}{l|}{\multirow{3}{*}{\begin{tabular}[c]{@{}l@{}}$x_4^T$, $x_4^F$\\ $x_5^T$, $x_5^F$\\ $x_6^T$, $x_6^F$\end{tabular}}}  & \multirow{3}{*}{\begin{tabular}[c]{@{}l@{}}$y_{1 \rightarrow 4}^F + x_4^T \rightarrow y_{1 \rightarrow 4}^F$\\ $y_5^F + x_5^T \rightarrow y_5^F$\\ $y_6^T + x_6^F \rightarrow y_6^T$\end{tabular}}  \\ \cline{1-4}
    \multicolumn{1}{|l|}{\multirow{3}{*}{\textbf{3}}} & \multirow{3}{*}{\textit{Add}}  & \multicolumn{1}{l|}{\multirow{3}{*}{\begin{tabular}[c]{@{}l@{}}$x_1^T$, $x_1^F$\\ $3 x_2^T$, $3 x_2^F$\\ $x_3^T$, $x_3^F$\end{tabular}}}  & \multicolumn{1}{l|}{\multirow{3}{*}{\begin{tabular}[c]{@{}l@{}}$y_1^F + x_1^T \rightarrow y_1^F$\\ $y_2^T + x_2^F \rightarrow y_2^T$\\ $y_3^T + x_3^F \rightarrow y_3^T$\end{tabular}}}  & \multicolumn{1}{l|}{} & \multicolumn{1}{c|}{}  & \multicolumn{1}{c}{}  & \multicolumn{1}{l|}{}  &  \\
    \multicolumn{1}{|l|}{}  &  & \multicolumn{1}{l|}{}  & \multicolumn{1}{l|}{}  & \multicolumn{1}{l|}{} & \multicolumn{1}{c|}{}  & \multicolumn{1}{c}{}  & \multicolumn{1}{l|}{}  &  \\ \cline{6-9} 
    \multicolumn{1}{|l|}{}  &  & \multicolumn{1}{l|}{}  & \multicolumn{1}{l|}{}  & \multicolumn{1}{l|}{} & \multicolumn{1}{l|}{\multirow{3}{*}{\textbf{10}}} & \multirow{3}{*}{\textit{Add}}  & \multicolumn{1}{l|}{\multirow{3}{*}{$d_y$}}  & \multirow{3}{*}{\begin{tabular}[c]{@{}l@{}}$y_{1 \rightarrow 4}^F + d_y \rightarrow d_y$\\ $y_5^F + d_y \rightarrow d_y$\\ $y_6^T + d_y \rightarrow d_y$\end{tabular}}  \\ \cline{1-4}
    \multicolumn{1}{|l|}{\multirow{3}{*}{\textbf{4}}} & \multirow{3}{*}{\textit{Add}}  & \multicolumn{1}{l|}{\multirow{3}{*}{$d_y$}}  & \multicolumn{1}{l|}{\multirow{3}{*}{\begin{tabular}[c]{@{}l@{}}$y_1^F + d_y \rightarrow d_y$\\ $y_2^T + d_y \rightarrow d_y$\\ $y_3^T + d_y \rightarrow d_y$\end{tabular}}}  & \multicolumn{1}{l|}{} & \multicolumn{1}{l|}{}  &  & \multicolumn{1}{l|}{}  &  \\
    \multicolumn{1}{|l|}{}  &  & \multicolumn{1}{l|}{}  & \multicolumn{1}{l|}{}  & \multicolumn{1}{l|}{} & \multicolumn{1}{l|}{}  &  & \multicolumn{1}{l|}{}  &  \\ \cline{6-9} 
    \multicolumn{1}{|l|}{}  &  & \multicolumn{1}{l|}{}  & \multicolumn{1}{l|}{}  & \multicolumn{1}{l|}{} & \multicolumn{1}{l|}{\textbf{11}}  & \textit{Add}  & \multicolumn{1}{l|}{$d_y$}  & $d_y + d_y \rightarrow \emptyset$  \\ \cline{1-4} \cline{6-9} 
    \multicolumn{1}{|l|}{\textbf{5}}  & \textit{Add}  & \multicolumn{1}{l|}{$d_y$}  & \multicolumn{1}{l|}{$d_y + d_y \rightarrow \emptyset$}  & \multicolumn{1}{l|}{} & \multicolumn{1}{l|}{\multirow{3}{*}{\textbf{12}}} & \multirow{3}{*}{\textit{Add}}  & \multicolumn{1}{l|}{\multirow{3}{*}{\begin{tabular}[c]{@{}l@{}}$y_{4 \rightarrow 7}^T$, $y_{5 \rightarrow 7}^T$\\ $y_{6 \rightarrow 7}^T$, $y_7^F$\end{tabular}}} & \multirow{3}{*}{\begin{tabular}[c]{@{}l@{}}$x_4^F + y_{4 \rightarrow 7}^T \rightarrow \emptyset$\\ $x_5^F + y_{5 \rightarrow 7}^T \rightarrow \emptyset$\\ $x_6^T + y_7^F \rightarrow \emptyset$\end{tabular}} \\ \cline{1-4}
    \multicolumn{1}{|l|}{\multirow{6}{*}{\textbf{6}}} & \multirow{6}{*}{\textit{Add}}  & \multicolumn{1}{l|}{\multirow{6}{*}{\begin{tabular}[c]{@{}l@{}}$y_4^T$, $y_{1 \rightarrow 4}^F$\\ $y_5^T$, $y_{2 \rightarrow 4}^F$\\ $y_5^F$, $y_{2 \rightarrow 6}^F$\\ $y_6^T$, $y_{3 \rightarrow 6}^F$\end{tabular}}} & \multicolumn{1}{l|}{\multirow{6}{*}{\begin{tabular}[c]{@{}l@{}}$x_1^F + y_4^T \rightarrow \emptyset$\\ $x_2^T + y_{2 \rightarrow 4}^F \rightarrow \emptyset$\\ $x_2^T + y_5^T \rightarrow \emptyset$\\ $x_2^T + y_{2 \rightarrow 6}^F \rightarrow \emptyset$\\ $x_3^T + y_{3 \rightarrow 6}^F \rightarrow \emptyset$\end{tabular}}} & \multicolumn{1}{l|}{} & \multicolumn{1}{l|}{}  &  & \multicolumn{1}{l|}{}  &  \\
    \multicolumn{1}{|l|}{}  &  & \multicolumn{1}{l|}{}  & \multicolumn{1}{l|}{}  & \multicolumn{1}{l|}{} & \multicolumn{1}{l|}{}  &  & \multicolumn{1}{l|}{}  &  \\ \cline{6-9} 
    \multicolumn{1}{|l|}{}  &  & \multicolumn{1}{l|}{}  & \multicolumn{1}{l|}{}  & \multicolumn{1}{l|}{} & \multicolumn{1}{c|}{\textbf{13}}  & \multicolumn{1}{c}{\textit{Add}}  & \multicolumn{1}{l|}{$d_x$}  & \textit{No Rules Apply}  \\ \cline{6-9} 
    \multicolumn{1}{|l|}{}  &  & \multicolumn{1}{l|}{}  & \multicolumn{1}{l|}{}  & \multicolumn{1}{l|}{} & \multicolumn{1}{c|}{\textbf{14}}  & \multicolumn{1}{c}{\textit{Add}}  & \multicolumn{1}{l|}{$d_x$}  & $d_x + d_x \rightarrow \emptyset$  \\ \cline{6-9} 
    \multicolumn{1}{|l|}{}  &  & \multicolumn{1}{l|}{}  & \multicolumn{1}{l|}{}  & \multicolumn{1}{l|}{} & \multicolumn{1}{c|}{\textbf{15}}  & \multicolumn{1}{c}{\textit{Add}}  & \multicolumn{1}{l|}{$x_7^T$, $x_7^F$}  & $y_{6 \rightarrow 7}^T + x_7^F \rightarrow y_{6 \rightarrow 7}^T$  \\ \cline{6-9} 
    \multicolumn{1}{|l|}{}  &  & \multicolumn{1}{l|}{}  & \multicolumn{1}{l|}{}  & \multicolumn{1}{c|}{} & \multicolumn{1}{l|}{\textbf{16}}  & \textit{Add}  & \multicolumn{1}{l|}{$d_y$}  & $y_{6 \rightarrow 7}^T + d_y \rightarrow d_y$  \\ \cline{1-4} \cline{6-9} 
    \multicolumn{1}{|c|}{\textbf{7}}  & \multicolumn{1}{c}{\textit{Add}} & \multicolumn{1}{l|}{$d_x$}  & \multicolumn{1}{l|}{\textit{No Rules Apply}}  & \multicolumn{1}{c|}{} & \multicolumn{1}{l|}{\textbf{17}}  & \textit{Add}  & \multicolumn{1}{l|}{$d_y$}  & $d_y + d_y \rightarrow \emptyset$  \\ \hline
    \end{tabular}

\caption{\textit{(2, 0) and (2, 1) rules and steps to compute the circuit in Figure \ref{fig:(2,0)_example} with Figure \ref{fig:aoc1}'s indexing.}}\label{tab:(2,0)_example}
\end{center}
\end{table}

With the computation of individual gates demonstrated in our system, we now expand these features to computing entire circuits. We begin with a simple example in Figure \ref{fig:(2,0)_example} to show the concepts before giving the general construction.

Our example circuit has three inputs: $x_1$, $x_2$, and $x_3$. In the first layer, $x_2$ is fanned out three times. One is fanned into an AND gate with $x_1$, another fanned into a NOT gate, and the other fanned into an AND gate with $x_3$. Finally, at the next depth level, the output of all three gates are fanned into an OR gate, whose output is the final circuit output.

Table \ref{tab:(2,0)_example} shows how to compute the circuit in Figure \ref{fig:(2,0)_example}. The primary inputs of the circuit in Figure \ref{fig:(2,0)_example} are represented by the species in the initial configuration. Steps 1-5 fan out the second primary input, convert the output species ($y_n^b$) into input species ($x_n^b$), and delete excess species. Step 6 computes the gates at the first depth level. Steps 7-11 convert the output species into input species and deletes excess species. Step 12 computes the final gate. Steps 13-17 delete excess species and converts the output of the final gate into an input species that represents the solution to the circuit ($x_7^T$).

\vspace{-.2cm}
\subsection{Computing Circuits with (2,0) Void and (2,1) Catalyst Rules}\label{2_0_catalysts}
\vspace{-.1cm}

\begin{theorem} \label{(2,0)(2,1)_circuit}
Threshold circuits (TC) can be strictly computed with $(2,0)$ void rules and $(2,1)$ catalyst rules, $O(G)$ species, $O(D)$ steps, and $O(W)$ volume.
\end{theorem}
\begin{proof}
With only $(2, 0)$ rules, there is no known way to perform fan-outs without introducing, at most, an exponential count of species at certain steps. This makes it impossible to strictly compute circuits with only $(2, 0)$ rules and results in a large volume. The use of $(2, 1)$ catalyst rules enables the step CRN to compute with arbitrary fan-out without an increase in species count, as well as deleting all species that are no longer needed. This allows for strict computation and decreases the volume of the step CRN to a polynomial size.

The initial configuration of this step CRN should consist of a $y_n^b$ species with the appropriate indexing and truth values for each primary input. Section \ref{2_0_gates} explains how to compute TC gates. To run the circuit, we must convert the output species into input species. In addition, this step CRN uses $d_x$ and $d_y$ as \emph{deleting} species. Table \ref{tab:cat_fan_out} shows the steps and rules needed to perform arbitrary fan-out for a gate. All outputs at the same depth level can be fanned out at the same time.

\begin{table}[t]
\vspace{-.2cm}
\renewcommand{\arraystretch}{.5}
    \begin{center}
    \begin{tabular}{| c | c  c | c | c |}\hline
    
    \multicolumn{3}{|c|}{\textbf{Steps}} & \textbf{Relevant Rules} & \textbf{Description} \\ \hline
    
    \textbf{1} & \textit{Add} & \makecell{$d_x$} 
    
    & \makecell{\vspace{-8pt} \\ $\forall n \in \{1,\cdots,G\}:$ \\ $\forall b \in \{T,F\}$ \\ $d_x + x_n^b \rightarrow d_x$ \\ $d_x + a_n^b \rightarrow d_x$ \\ $d_x + b_n^b \rightarrow d_x$ \\ \vspace{-8pt}}
    
    & \makecell{Delete all input species ($x_n^b$) and helper \\ species that are no longer needed.} \\ \hline

    \textbf{2} & \textit{Add} & \makecell{$d_x$} 
    
    & \makecell{\vspace{-8pt} \\ $d_x + d_x \rightarrow \emptyset$ \\ \vspace{-8pt}}
    
    & \makecell{Remove deleting species $d_x$.} \\ \hline

    \textbf{3} & \textit{Add} & \makecell{$|f_i^{out}| \cdot x_i^T$ \\ $|f_i^{out}| \cdot x_i^F$} 
    
    & \makecell{\vspace{-8pt} \\ $y_i^T + x_i^F \rightarrow y_i^T$ \\ $y_i^F + x_i^T \rightarrow y_i^F$ \\ $\forall j \in f_i^{in}:$ \\ $y_{j \rightarrow i}^T + x_i^F \rightarrow y_{j \rightarrow i}^T$ \\ $y_{j \rightarrow i}^F + x_i^T \rightarrow y_{j \rightarrow i}^F$ \\ \vspace{-8pt}}
    
    & \makecell{Add species representing true and false \\ inputs and delete the species that are the \\ complement of the output. A single output \\ species can assign the truth value for as \\ many input species as needed.} \\ \hline

    \textbf{4} & \textit{Add} & \makecell{$d_y$} 
    
    & \makecell{\vspace{-8pt} \\ $\forall n \in \{1,\cdots,G\}:$ \\ $d_y + y_n^T \rightarrow d_y$ \\ $d_y + y_n^F \rightarrow d_y$ \\ $\forall j \in f_i^{in}:$ \\ $d_y + y_{j \rightarrow i}^T \rightarrow d_y$ \\ $d_y + y_{j \rightarrow i}^F \rightarrow d_y$ \\ \vspace{-8pt}}
    
    & \makecell{Delete all output species ($y_n^b$) that no \\ longer needed.} \\ \hline

    \textbf{5} & \textit{Add} & \makecell{$d_y$} 
    
    & \makecell{\vspace{-8pt} \\ $d_y + d_y \rightarrow \emptyset$ \\ \vspace{-8pt}}
    
    & \makecell{Remove deleting species $d_y$.} \\ \hline
    
    \end{tabular}
    
    \caption{\textit{(2, 0) and (2, 1) rules and steps for a gate with arbitrary fan out.}}\label{tab:cat_fan_out}
\end{center}
\end{table}

\textbf{Complexity:} Having a constant amount of unique species represent each gate in a circuit and a constant number of helper species results in $O(G)$ species.
All gates at a given depth level are computed at the same time in a constant number of steps. Thus, the circuit simulation requires $O(D)$ steps.
This step CRN only needs to introduce a constant number of species to compute each gate, and it deletes all species no longer needed after computing all gates at a given depth level. Thus, the step CRN  requires $O(W)$ volume.
\end{proof}

\subsection{Computing Circuits with (2,1) Catalyst Rules}\label{2_1_catalysts}

It's worth noting that (2,1) catalyst rules are able to compute TCs on their own. The main drawback is that there is no known way to directly compute majority gates without $(k \geq 2,0)$ void rules. Thus, any majority gate must be computed using AND, OR, and NOT gates when using only catalyst rules. Furthermore, deleting species that are no longer needed is slightly more convoluted with (2,1) rules compared to pure void rules. 

\begin{corollary} \label{(2,1)_circuit}
Threshold circuits (TC) can be strictly computed with only $(2,1)$ catalyst rules, $O(G)$ species, $O(D \log F_{maj})$ steps, and $O(W)$ volume.
\end{corollary}
\begin{proof}
Section \ref{2_0_gates} explains how to compute AND/OR/NOT gates using (2,0) rules, and they can easily be changed to (2,1) rules by making the $x_j^b$ species a catalyst. The method used to perform arbitrary fan out in Theorem \ref{(2,0)(2,1)_circuit} can be slightly modified to function with only (2,1) rules. Table \ref{tab:(2,1)_fan_out} demonstrates how this can be done. A special property of using (2,1) rules to compute gates is that the counts of the species being added are flexible. This is not the case when gates are computed with pure void rules, as it is necessary for the counts of certain species to be precise. For example, while Step 3 in Table \ref{tab:cat_fan_out} and Step 5 in Table \ref{tab:(2,1)_fan_out} are functionally equivalent steps, they have different computing requirements. When computing with (2,0) rules, we need exactly $|f_i^{out}|$ amount of $x_i^b$ species by the end of that step. On the other hand, when computing with (2,1) rules, we only need one copy of $x_i^b$, and even if multiple copies of that species were added, it would not have a significant impact on the computation of the circuit.

\textbf{Complexity:} The techniques used to compute circuits are functionally equivalent to the ones used in Theorem \ref{(2,0)(2,1)_circuit}, so the upper bound of species and volume remain the same, that is, $O(G)$ and $O(W)$, respectively.

Since majority gates must be computed using AND and OR gates when using only (2,1) rules, the depth of the circuit must increase. The conversion of a majority gate to AND/OR/NOT gates can be achieved with $O(n)$ gates and $O(\log n)$ depth where $n$ is the number of input bits of the majority gate \cite{sergeev2014upper} (any symmetric boolean function has a circuit of depth $O(\log n)$ and size $O(n)$ for $n$ bits). Thus, the maximum number of steps needed to compute a circuit would be $O(D \log F_{maj})$, where $F_{maj}$ is the maximum fan-in of any majority gate in the circuit.
\end{proof}

\begin{table}[t]
\vspace{-.2cm}
\renewcommand{\arraystretch}{.5}
    \begin{center}
    \begin{tabular}{| c | c  c | c | c |}\hline
    
    \multicolumn{3}{|c|}{\textbf{Steps}} & \textbf{Relevant Rules} & \textbf{Description} \\ \hline

    \textbf{1} & \textit{Add} & \makecell{$d_x'$} 
    
    & \makecell{\vspace{4pt} \\ $d_x' + d_x'' \rightarrow d_x'$ \\ \vspace{4pt}}
    
    & \makecell{Deleting species $d_x''$ makes it possible for \\ species $d_x$ to exist in the next step \\ without complications.} \\ \hline

    \textbf{2} & \textit{Add} & \makecell{$d_x$} 
    
    & \makecell{\vspace{-8pt} \\ $d_x + d_x' \rightarrow d_x$ \\ $\forall n \in \{1,\cdots,G\}:$ \\ $\forall b \in \{T,F\}$ \\ $d_x + x_n^b \rightarrow d_x$ \\ $d_x + a_n^b \rightarrow d_x$ \\ $d_x + b_n^b \rightarrow d_x$ \\ \vspace{-8pt}}
    
    & \makecell{Deleting species $d_x'$ makes it possible for \\ species $d_x''$ to exist in the next step \\ without complications. \\ Delete all input species ($x_n^b$) and helper \\ species that are no longer needed.} \\ \hline

    \textbf{3} & \textit{Add} & \makecell{$d_x''$} 
    
    & \makecell{\vspace{-8pt} \\ $d_x + d_x'' \rightarrow d_x''$ \\ \vspace{-8pt}}
    
    & \makecell{Removes deleting species $d_x$.} \\ \hline

    \textbf{4} & \textit{Add} & \makecell{$d_y'$} 
    
    & \makecell{\vspace{4pt} \\ $d_y' + d_y'' \rightarrow d_y'$ \\ \vspace{4pt}}
    
    & \makecell{Deleting species $d_y''$ makes it possible for \\ species $d_y$ to exist in the next step \\ without complications.} \\ \hline

    \textbf{5} & \textit{Add} & \makecell{$x_i^T$ \\ $x_i^F$} 
    
    & \makecell{\vspace{-8pt} \\ $y_i^T + x_i^F \rightarrow y_i^T$ \\ $y_i^F + x_i^T \rightarrow y_i^F$ \\ $\forall j \in f_i^{in}:$ \\ $y_{j \rightarrow i}^T + x_i^F \rightarrow y_{j \rightarrow i}^T$ \\ $y_{j \rightarrow i}^F + x_i^T \rightarrow y_{j \rightarrow i}^F$ \\ \vspace{-8pt}}
    
    & \makecell{Add species representing true and false \\ inputs and delete the species that are the \\ complement of the output. A single output \\ species can assign the truth value for as \\ many input species as needed.} \\ \hline

    \textbf{6} & \textit{Add} & \makecell{$d_y$} 
    
    & \makecell{\vspace{-8pt} \\ $d_y + d_y' \rightarrow d_y$ \\ $\forall n \in \{1,\cdots,G\}:$ \\ $d_y + y_n^T \rightarrow d_y$ \\ $d_y + y_n^F \rightarrow d_y$ \\ $\forall j \in f_i^{in}:$ \\ $d_y + y_{j \rightarrow i}^T \rightarrow d_y$ \\ $d_y + y_{j \rightarrow i}^F \rightarrow d_y$ \\ \vspace{-8pt}}
    
    & \makecell{Deleting species $d_y'$ makes it possible for \\ species $d_y''$ to exist in the next step \\ without complications. \\ Delete all output species ($y_n^b$) that are \\ no longer needed.} \\ \hline

    \textbf{7} & \textit{Add} & \makecell{$d_y''$} 
    
    & \makecell{\vspace{-8pt} \\ $d_y + d_y'' \rightarrow d_y''$ \\ \vspace{-8pt}}
    
    & \makecell{Remove deleting species $d_y$.} \\ \hline
    
    \end{tabular}
    
    \caption{\textit{(2, 1) rules and steps for a gate with arbitrary fan out.}}\label{tab:(2,1)_fan_out}
\end{center}
\vspace{-1.3cm}
\end{table}

\vspace{-.5cm}
\section{Lower Bounds and Hardness}\label{sec:LBs}
\vspace{-.2cm}

In this section, we prove negative results for computing with step CRNs. First, we show there exists a family of functions that require a logarithmic number of steps to compute. Then, we show hardness of verifying whether a step CRN properly computes a given function.

\vspace{-.2cm}
\subsection{Step Lower Bound for Controlled NOT}
\vspace{-.2cm}

\para{CNOT.} The Controlled NOT gate is a 2-bit input and 2-bit output gate taking inputs $X$ and $Y$, and outputting $X$ and $X \oplus Y$. In other words, the gate flips $Y$ if $X$ is true.

\para{k-CNOT.} We generalize this to a Controlled $k$-NOT gate. This is a $(k + 1)$-bit gate with inputs $X, Y_1, \ldots , Y_k$. The $Y$ bits all flip if $X$ is true. 
We choose this function since it has the property that changing $1$ bit of the input changes a large number of output bits. 


\para{Configuration Distance.} Recall configurations are defined as vectors. For two configurations $c_0, c_1$, we say the distance between them is $||c_0 - c_1||_1$, i.e., the sum of the absolute value of each entry in $c_0 - c_1$. 

\begin{lemma}\label{lem:ccnotSignal} Let $r$ be a positive integer parameter. 
    For all step CRNs $\Gamma$ with void rules of size $(r_1, 0)$ with $r_1\le r$ and pairs of initial configurations $c_T$ and $c_F$ with distance $2$ and equal volume, for any configuration $c_{Ts}$ terminal in the step $s$ from $c_T$, there exists a configuration $c_{Fs}$ terminal in step $s$ from $c_F$ such that the distance between $c_{Ts}$ and $c_{Fs}$ is ${O}(r^s)$.
\end{lemma}
\begin{proof}

     Let $T$ be the species that is in configuration $c_T$ and not in $c_F$. Similarly, define the species $F$ to be the species in configuration $c_F$ and not in $c_T$. 
     Consider the reaction sequence $R_T$ starting from $c_T$ and ending in $c_{T1}$. All but one reaction in $R_T$ can be applied to $c_F$. 
     Consider the reaction sequence $R_F$ that differs from $R_T$ by two reactions. The first is the reaction in $R_T$ that consumes $T$ and $r-1$ other species, and the second is a reaction that consumes $F$ and $r-1$ other species. Applying $R_F$ results in a configuration $c_{F1}$ that differs from $c_{T1}$ by at most $2r$ species.

     Now, assume there are two initial configurations in step $s$ with distance $2r^{s-1}$ away from each other. 
     In the base case, each different species between the configurations can be used in at most one rule, thus the species can only propagate $r-1$ additional changes in the next terminal configuration. 
     This results in a difference of $2rr^{s-1}  = 2r^s$ in the $s$ stage. 
\end{proof}

The configuration distance between two output configurations is related to the Hamming distance of the output strings they represent. Lemma \ref{lem:ccnotSignal} can be used to get a get a logarithmic lower bound for the number of steps required when we fix our rule size to be a constant. 

\begin{theorem}\label{thm:ccnotLB}
    For all constants $r$, any CRN that strictly computes a $k$-CNOT gate with rules of size $(r_1, 0)$ satisfying $r_1\le r$ requires $\Omega(\log k)$ steps.
\end{theorem}
\begin{proof}
    Flipping only the $X$ bit of the the input changes all $k+1$ output bits. It follows that in order to  compute a $k$-CNOT, we must have at least $k$ distance between the two final configurations even when starting from configurations with distance $1$. We can also assume these have the same volume since by changing a bit value we are only changing which species we add. With Lemma \ref{lem:ccnotSignal}, we get the following inequality and must compute $s$.
    \vspace{-.3cm}
    \[k \leq 2r^s \hspace{1cm}\rightarrow \hspace{1cm} \log k \leq 2s \log r \hspace{1cm}\rightarrow \hspace{1cm} \frac{\log k}{2\log r} \leq s \]
    
    \vspace{-.2cm}
    \noindent Since $r$ is a constant we get an asymptotic bound of    
    $s=\Omega(\log k)$.
\end{proof}

We also note the $k$-CNOT can be computed by $k$ XOR gates in parallel. This implies this lower bound does not hold with catalytic reactions either as Theorem \ref{(2,0)(2,1)_circuit} shows this can be computed in $O(1)$ steps or without the input-strict requirement. This is because increasing the fan-out of the $X$ bit does not incur a cost in the number of steps in both of these generalizations. Plugging this XOR circuit into Theorem \ref{(3,0)_circuit} gives a bound of $\Theta(\log k)$ steps showing the construction is optimal for some circuits. 

\vspace{-.2cm}
\subsection{Function Verification Hardness}
\vspace{-.2cm}

We have established that void step CRNs can simulate Boolean circuits.  We now discuss the complexity of the computational problem of determining if a given (void) step CRN does compute a given function.  Specifically, we consider the following decision problem: 

\textbf{(Strict Function Verification):} Given a step CRN $C_S=(S, X, Y)$ and a Boolean function $f(\cdot)$\footnote{We assume that $f(\cdot)$ is given in the form of a circuit $c_f$. We leave as future work the complexity of other representations such as a truth table. } where $f(x_1,\cdots, x_n)=y_1:\{0,1\}^n\rightarrow \{0,1\}$, decide if $C_S$ computes Boolean function $f(\cdot)$. In particular, let $f_0(x_1,\cdots, x_n)=false$, which is false for all inputs.

Theorem~\ref{coNP-hard-thm} shows that strict function verification in a step CRN system with void rules is coNP-hard, and coNP membership for this problem is shown in Theorem \ref{coNP-m}.

\begin{theorem}\label{coNP-hard-thm}
It is coNP-hard to determine if a  given a $\mathcal{O}(1)$-step CRN $C_S=(S, X, Y)$ with $(3,0)$ rules computes the boolean function $f_0(x_1,\cdots, x_n)$.
\end{theorem}

\vspace{-.2cm}

\begin{proof} The decision problem 3SAT is a classical NP-complete problem. Its complementary problem, which is coNP-complete, is to decide if the conjunction of several clauses with three literals cannot be satisfied. We also reduce from the special case of 3SAT where each variable appears at most $4$ times, shown to be NP-complete in~\cite{tovey1984simplified}.
Let $F(x_1,\cdots, x_n)=C_1\wedge C_2\wedge\cdots C_m$ 
be an instance for a 3SAT problem, where each $C_i$ is a clause that has at most three literals, for example, $C_1=(x_1\vee \overline{x_2}\vee x_3)$. The function  $F(.)$ can be computed by a boolean circuit of constant depth. Checking if $F(x_1,\cdots, x_n)=f_0(x_1,\cdots, x_n)$ for all $(x_1,\cdots, x_n)$ is the complementary problem for 3SAT. 

It follows from Theorem~\ref{(3,0)_circuit}, which shows $F(.)$ can be simulated by a step CRN $C_S=(S,X,Y)$ with $(3,0)$ rules. The number of steps of the CRN is $\mathcal{O}(D\log F_out)$. The depth of the circuit is constant since all clauses can be computed in parallel and the gates handle arbitary fan-in. The fanout is also constant because each variable only appears $4$ times.  
\end{proof}

\vspace{-.2cm}

\begin{theorem}\label{coNP-m}
Determining if a  given a $s$-step CRN $C_S=(S, X, Y)$ with $(r,0)$ rules computes the boolean function $f_0(x_1,\cdots, x_n)$ is in coNP. 
\end{theorem}

\vspace{-.2cm}
\begin{proof}
This problem can be solved by a polynomial time non-deterministic algorithm which does the following. Pick a string $b$ of length $n$ and compute $f_0(b) = Y$. Then convert $b$ to the initial configuration $X(b)$. Guess a sequence of $s$ terminal configurations $c_1, \ldots, c_s$ where $c_i$ is a terminal configuration in the $i$th step. To verify this, call the NP algorithm for reachability in Volume Decreasing CRNs from \cite{Alaniz:2022:ARXIV} to verify each configuration is reachable in the correct step. If the final configuration $c_s$ does not represent $Y$ then reject. 
\end{proof}

\vspace{-.2cm}
\begin{theorem}\label{coNP-c}
It is coNP-Complete to determine if a  given a $O(1)$-step CRN $C_S=(S, X, Y)$ with $(3,0)$ rules computes the boolean function $f_0(x_1,\cdots, x_n)$. 
\end{theorem}
\vspace{-.2cm}
\begin{proof}
    Follows from Theorems \ref{coNP-hard-thm} and \ref{coNP-m}.
\end{proof}
\vspace{-.4cm}
\section{Conclusions and Future Work}\label{sec:conclusion}
\vspace{-.2cm}

We have proposed the step CRN model, a natural augmentation to the CRN model, and shown that void rule CRNs, a simple but computationally weak class of CRNs, become capable of efficiently simulating Threshold Circuits under this extension.  We have shown this holds even when limited to $(3,0)$ reactions, and further shown that bi-molecular reactions are equally powerful if permitted access to catalytic reactions.  We also show that the step augmentation is fundamentally needed:  without access to a super-constant number of steps, such computation is impossible.  Finally, we utilize our positive results to show coNP-completeness for the problem of deciding if a given step CRN computes a given function.

The step CRN model presented in this work, along with our results, naturally lead to a number of additional promising research directions.  A small sample of these are:
\begin{itemize}
    \item Lower Bounds and Catalysts: We conjecture $(3,0)$ void rules are the smallest size rules for strictly simulating TC without the use of a catalyst. 
    Further, we show that more than a constant number of steps are required for circuit computation for non-catalytic void rules, but is it true with a catalyst?

    
    \item Robustness:  While void rules potentially offer a simpler path to experimental feasibility and scalability, some of our techniques require precise counts of species to be added at different steps of the computational process.  Such precision is a hurdle to experimental implementation.  Thus, it is interesting to consider to what extent these results can be made robust to  approximate counts (or have a lower-bound). 
    
    \item Reachability: The \emph{reachability} problem of determining if a given configuration is reachable from an initial configuration is well-studied in CRNs and other computational models. We showed that steps allow for greater computational power with void rules. How does the addition of steps affect the reachability problem, and specifically, what is the complexity when relating void rules, catalysts, rule size, and number of steps?  
    


    \item General Staged CRNs:  We explored a simple scheme for including separate stages into the CRN model by simply adding new species at each step.  A more general modelling could include multiple separate bins that may be mixed or split together over a sequence of stages.  Formalizing such a model and exploring the added power of this generalization is an interesting direction for future work.

\end{itemize}



\bibliography{crn}


\end{document}